\documentstyle[osa,manuscript]{revtex}  
\begin{document}                
\title{Right-unitary transformation theory and applications}
\author{Zhong ~Tang* }

\address{School of Physics, Georgia Institute of Technology, Atlanta, 
Georgia 30332-0430}

\maketitle
\begin{abstract}
We develop a new transformation theory in quantum physics, where 
the 
transformation operators, defined in the infinite dimensional 
Hilbert space, have right-unitary inverses only. 
Through several theorems, we discuss 
the  properties of state space of such operators.
As one application of the right-unitary transformation (RUT),
we show  that using 
the RUT method, we can solve exactly various interactions of 
many-level atoms with quantized radiation fields, 
where the energy of atoms can be two levels, three levels 
in $\Lambda, ~V$ and $\equiv$ configurations, and up to
higher ($>3$) levels. These interactions have wide applications
in atomic physics, quantum optics and quantum electronics.
In this paper, we focus on two typical systems:
one is a two-level
generalized Jaynes-Cummings model, where the cavity field varies 
with the external source; the other one is the interaction
of three-level
atom with quantized radiation fields, where the atoms 
have $\Lambda$-configuration energy levels, and the radiation fields 
are
one-mode or two-mode cavities.\\

\noindent PACS number(s): 03.65.-w, 42.50.-p\\

\noindent *Electronic address: zt2@prism.gatech.edu
          
\end{abstract}

\newpage
\begin{center}
\section*{I. Introduction}               
\end{center}

Not only an important method, 
but also an intrinsic description of symmetry
for the physical systems,  transformation theory is always an 
interesting topic in physics that  has acquired a lot of studies.  
The  transformations preserving the measurements are called 
$physical$ transformations. 
We know that classical mechanics is founded on 
the symplectic manifold, where the transformations of momentum 
and position preserving the symplectomorphism are 
the physical transformations,
which form a group with composition, called the symplectic group.
These transformations are also called $canonical$ transformations,
since they leave both the canonical equations of motion 
and the Poisson brackets invariant.

Along the jump from classical theory to quantum theory,
the Poisson brackets are replaced by the bosonic commutation
relations. The linear transformations of bosons those leave
the bosonic commutation relations invariant were first introduced by
Bogoliubov \cite{bogo}. These bosonic transformations were found to 
form the same symplectic group as that 
in classical mechanics, even though these two kinds of 
transformations have different physical meanings essentially
 \cite{moskin}.
The quantum canonical transformations of fermions  were 
first introduced by   Valatin \cite{vala}
in his study of superconductivity. Since fermions are the purely 
quantum objects without classical correspondence,
the Lie group formed by the fermionic transformations is no longer
the symplectic group, but a sub-Lie group isomorphic with the
$rotation$ group \cite{zhang}. The Bogoliubov-Valatin
transformations as the  linear quantum transformations have
wide applications, for examples in the BCS theory \cite{vala} of
superconductivity and in the calculation 
of blackhole radiation \cite{hwk}.   Recently, 
the supersymmetric transformation that mixes 
the boson fields with the fermion fields in a unified form advances 
the transformation 
theory to a big stage \cite{super}. 
However,  some  interesting
physical predictions raised by the supersymmetric theory
are still awaiting  experiment to test.
In quantum gauge theory,
the BRST transformations of the gauge fields 
and the ghost fields  are
in the well-known supersymmetric forms \cite{brst}, 
which play an important 
role in the renormalization proof of the Standard Model. 
In sum, the quantum transformation theory 
is still under development
for various purposes.
We notice that the above transformations are the 
$unitary$ transformations. Since quantum theory is set upon 
the Hilbert space,   and 
the duality of the Hilbert space is  defined through
the Hermitian conjugate, it has been recognized that only the 
unitary transformations do not change the Hilbert space.
On the other hand, the nonunitary transformations can not preserve 
the Hermitian duality, 
thus  break the realities of physical observables and probabilities. 
That is why the nonunitary transformations 
are always rejected in quantum theory \cite{time}.

In a previous  paper \cite{tang}, we introduced an alternative 
method: 
right-unitary transformation (RUT), to deal with
the two-level Jaynes-Cummings model \cite{jcm}, which is
a basis of the fully-quantum
description of radiation-matter interaction, and widely used in
quantum optics, quantum electronics, etc. It was defined
in Ref. [9] that if an operator $U$
satisfies the conditions: $ UU^{\dagger}=1,~ U^{\dagger}U\neq 1$,
it belongs to RUT. In a strict sense,
$U$ is a special nonunitary operator that has  
a right-unitary inverse only. In the matrix representation,
$U$ can only be the matrix in an infinite dimensional space. 
Such an operator has been recognized 
as the $irregular$ operator by mathematician, thus obtains few 
studies,
let alone its utility in physics.
However, in Ref. [9], we  found that 
various JC models can be solved exactly by the RUT method.
This method not only shows its own merits 
such as simplicity and general applicability,
but also leads to a deep
understanding of the JC models in essence.
This work further implies that the transformations 
applicable to quantum systems should not be restricted 
in the unitary range. Instead, some nonunitary transformations
which were regarded as the irregular objects might have their 
particular utilities in physics.

We know that the unitary transformations do not change the 
measurements
of a quantum system.
But the situation is quite different when using
the nonunitary transformations. 
In this paper, we attempt to develop 
a theory of the right-unitary transformation in mathematics, and 
discuss its applications in physics.
 
This paper is arranged as follows.
In Section {\bf II}, through six theorems, we reveal
some common properties of the right-unitary transformation (RUT),
and conclude a general procedure on how to apply 
RUT to quantum systems.
Section {\bf II} is the foundation of the whole work, it includes 
three subsections.
In Subsection {\bf 2.1}, we  reveal that the state space of
any operator $U\in$ RUT is composed of two independent
subspaces $\{~|\Psi^{0}\rangle~\}$ and $\{~|\Psi^{1}\rangle~\}$,
where  $\{~|\Psi^{0}\rangle~\}$ is called the $Kernel$ of $U$, which 
usually has a finite number of basic elements, and 
satisfies the equation: $U\{~|\Psi^{0}\rangle~\}=0$. On the other hand, 
in the subspace $\{~|\Psi^{1}\rangle~\}$, 
$U$ acts as a unitary operator. 
Similar to the unitary transformations, 
all the operator $U\in$ RUT in the same Hilbert space
form a semigroup, which is called right-unitary semigroup.
For a certain system with Hamiltonian $H$, through a theorem we 
show
 that if the Kernel of $U$ is checked to be 
isomorphic with a subset of the eigenkets of $H$, 
$U$ can be used as a unitary
transformation to the remaining subspace of $H$
without changing the spectrum. 
In Subsection {\bf 2.2}, we briefly discuss the application of RUT
to the nonstationary system.
In order to construct the RUT conveniently in the Fock space, 
 we employ the recently
introduced inverses ( $b, ~ b^{\dagger} $) 
of boson creation and annihilation operators ($ a^{\dagger},~a $)
\cite{arvind} in Subsection {\bf 2.3}, and show that the well-known
quantum phase operators \cite{phasor} constructed by $ a$ and 
$a^{\dagger} $
(or $b$ and $ b^{\dagger}$) form an Abelian subgroup of the  
right-unitary semigroup. 

Section {\bf  III} attributes to one of the applications
of RUT in physics, where we use the RUT method to treat the systems 
of many-level atoms interacting 
with the quantized radiation fields, where the RUT
are constructed by the quantum phase operators.
This section contains four subsections.
In Subsection {\bf  3.1}, we approach such a model that the
atoms have two energy levels, and the radiation field is 
designed to be a one-mode cavity that varies with 
time. Through this model we hope to
achieve the goal of controlling the effect such 
as atomic inversion of the system $via$
the external source. This model can be regarded as 
the nonstationary JC model, which is found to exhibit
such a property that there exist a particular relation between
the atomic inversion and the energy 
exchange of the atomic system with the external source.
Therefore, by measuring the energy exchange,
we can understand  the dependence of quantities such as 
the atomic inversion, and
the mean photon number on time (on the external source). 

In recent years, the case of three-level atoms 
interacting with quantized 
electromagnetic field have obtained extensive studies. 
The model where the atoms have $\Lambda$-configuration
energy level has been applied to a number of different 
contexts such as the coherent population trapping, 
laser cooling \cite{trap}, lasing without inversion \cite{inver},
and electromagnetically induced transparency (EIT) \cite{EIT}.
In Subsection {\bf  3.2},
 we use the RUT method to solve exactly
the atom-radiation interaction,
where the atoms have $\Lambda$-configuration energy levels, 
and the radiation field is designed to be a one-mode cavity field.
The case of a two-mode cavity is treated in Subsection {\bf  3.3}, 
which is further divided into two parts:
In the first part {\bf  3.3.1}, we consider such a situation
that the detunings of the system are enough large, 
then we are able to construct a unitary transformation 
to separate the upper level 
from the system, and 
the remaining two low levels are still treated by the RUT
method. In the second part {\bf  3.3.2}, we use the RUT method to 
solve the system exactly.
In Subsection {\bf  3.4}, we briefly show how to construct RUT 
to solve the matter-radiation interactions where 
the atoms can be three levels in $V$ or $\equiv$ configuration, and 
up to 
four levels. 
Section {\bf  IV} is the conclusion.

In the present paper, we focus on the RUT method, 
the analyses of the quantities   
related to the interesting phenomena such as lasing without 
inversion, and electromagnetically induced transparency 
(EIT) are  not given.
Since there are lots of formula and 
quantities in the paper, 
the meanings of the symbols are independent in each section.\\

\begin{center}
\section*{II. Right-unitary transformation}
\end{center}

\begin{center}
\subsection*{2.1.  Stationary case}
\end{center}

Consider a Hermitian operator $\Lambda$  with a discrete spectrum
\begin{equation}
\Lambda|\psi_{n}\rangle=\lambda_{n}|\psi_{n}\rangle,~~~~n=1,2,\cdots,k,
\end{equation} 
the eigenkets $|\psi_{n}\rangle$ are orthogonal mutually,
$\langle\psi_{m}|\psi_{n}\rangle=\delta_{mn}$. 
Let us choose an operator $U$ to transform  $\Lambda$  into
another frame $\Lambda'$. If $\Lambda'$ is still a Hermitian 
quantity
and has the same spectrum as
$\Lambda$, we call $U$ an $applicable ~transformation$ to 
$\Lambda$. There raise two questions subsequently: 

How  to transform $\Lambda$ into $\Lambda'$ by using $U$?
 
Since we require that $\Lambda'$  still be
a Hermitian quantity,  the transformation from  
$\Lambda$ to $\Lambda'$ is usually chosen as
$\Lambda'=U\Lambda U^{\dagger}$. Then

What is the basic requirement to $U$? 

This question is indeed the crux of the 
transformation theory, and has not a complete answer yet. 
Certainly, as we have pointed out in the 
Introduction that the unitary transformation
is the $applicable ~transformation$.
Here we  would like to see how far we can 
approach beyond the unitary transformation. 
As it required that $\Lambda'$ should have
the same spectrum as $\Lambda$, it is proper to                                                                          
choose an extreme case,
$\Lambda=1$, as an invariant in the  transformation.
This will result in such a requirement to $U$  
that $UU^{\dagger}=I$, in order 
to maintain the case of unity: 
$\Lambda=\Lambda'=1$.  We know that in the infinite dimensional
Hilbert space,  $UU^{\dagger}=I$ does not mean 
that $U$ is a unitary
operator. In fact, there exists the 
the following nonunitary transformation,
\begin{equation}
\left\{
\begin{array}{l}
UU^{\dagger}=I, \\
U^{\dagger}U=W\neq I.
\end{array}
\right.
\end{equation}
The aim of present work is to understand how this kind of
nonunitary transformation works in quantum theory.
We here call the operator $U$ satisfying Eq. (2)
right-unitary transformation (RUT), in order to distinguish it
from the other nonunitary transformations. We know that
RUT are the operators in the infinite 
dimensional space,
they can only be applied  to the system 
having the same infinite dimensional state space as RUT,
i.e., $k$ in Eq. (1) goes to infinity.

Before carrying a theoretical study on RUT, we first 
give a simple example of it: 
 quantum phase operators \cite{phasor},
\begin{equation}
F_{1}=\frac{1}{\sqrt{a a^{\dagger}}}a,~~~
F_{1}^{\dagger}=a^{\dagger}\frac{1}{\sqrt{a a^{\dagger} }},
\end{equation}
where $a$ and $a^{\dagger}$ are bosonic annihilation and creation
operators, respectively. It is easily calculated that
\begin{equation}
\left\{
\begin{array}{l}
 F_{1}F_{1}^{\dagger}=I,\\
F_{1}^{\dagger}~ F_{1}=
I-|0\rangle\langle 0|.
\end{array}
\right.
\end{equation}
Which indicate that $F_{1}$ belongs to the right-unitary operators. 
This concrete example of RUT
will be  helpful to understand the following mathematical approach. 
The properties of the quantum phase operators will be 
carefully discussed Subsection {\bf 2.3}.
 
We now study the properties of RUT through 
the following several theorems.

$ Theorem~I$~~~For any  operator $U\in$ RUT, 
that  $W\equiv U^{\dagger}U $ has a complete set 
of eigenkets $\{~|\Psi\rangle~\}$,
where  $\{~|\Psi\rangle~\}$  is constituted by two independent 
subsets: 
$\{~|\Psi^{0}\rangle~\}$ and $\{~|\Psi^{1}\rangle~\}$.
For the set $\{~|\Psi^{0}\rangle~\}$, 
$U$ acts as an annihilation operator,
\begin{equation}
U|\Psi^{0}_{i}\rangle=0, ~~
|\Psi^{0}_{i}\rangle\in \{~|\Psi^{0}\rangle~\};
\end{equation}
For the other set $\{~|\Psi^{1}\rangle~\}$, 
$U$ acts as a unitary operator, namely,
\begin{equation}
W|\Psi^{1}_{j}\rangle=|\Psi^{1}_{j}\rangle, 
~~
|\Psi^{1}_{j}\rangle\in \{~|\Psi^{1}\rangle~\};
\end{equation}

$Proof$:~~For any $U\in$ RUT, it follows from Eq. (2) that
\begin{equation} 
U(1-W)=0.
\end{equation}
Equation (7) is further left-multiplied by $U^{\dagger}$,  we have
\begin{equation}
W-W^{2}=0.
\end{equation}
Equation (8) indicates that the Hermitian operator $W$ has   
eigenvalues $0$ and $1$ only, and the corresponding eigenstates,
denoted by  $\{~|\Psi^{0}\rangle~\}$
and  $\{~|\Psi^{1}\rangle~\}$, form a complete set,
where the brackets are used to represent the case of 
degeneracy if exists. For any 
$|\Psi^{0}_{i}\rangle\in \{~|\Psi^{0}\rangle~\}$, using Eq. (7), 
we obtain $U(1-W)|\Psi^{0}_{i}\rangle=U|\Psi^{0}_{i}\rangle=0$, 
which proves Eq. (5). On the other hand,
for any $|\Psi^{1}_{j}\rangle\in \{~|\Psi^{1}\rangle~\}$, 
we have
$U^{\dagger}U|\Psi^{1}_{j}\rangle=W|\Psi^{1}_{j}\rangle=
|\Psi^{1}_{j}\rangle$, namely,
$U$ acts as a unitary operator in the subspace 
$\{~|\Psi^{1}\rangle~\}$.~\#

As the start point of our work, the above theorem reveals 
the general structure of the state space of RUT.
Applying this theorem to the
phase operator  $F_{1}$, we have:
$\{~|\Psi^{0}\rangle~\}=\{~|0\rangle~\}$, and 
$\{~|\Psi^{1}\rangle~\}=\{~|k\rangle,~ k=1,2,\cdots\infty~\}$.

The following two corollaries  are directly
associated with   Theorem I:

First, from Eq. (5), we know that
 
$Corollary~ I. a.$ ~~ For any operator $U\in$ RUT, there  
exists at least 
one vector $|\psi\rangle\neq 0$ to satisfies the equation: 
$U|\psi\rangle=0$.~\#

Second, for an arbitrary vector  $|\theta\rangle$, from  that
\begin{equation}
U^{\dagger}U|\theta\rangle=U^{\dagger}U\left[|\Psi^{0}\rangle
\langle\Psi^{0}|\theta\rangle
+|\Psi^{1}\rangle\langle\Psi^{1}|\theta\rangle
\right]=\langle\Psi^{1}|\theta\rangle
|\Psi^{1}\rangle,
\end{equation}
 (the brackets are omitted here) we obtain:
 
$Corollary~I. b$: ~~For  $U\in$ RUT, $U^{\dagger}U$ is 
a projection operator that
maps an arbitrary vector into the unitary subspace of $U$.~\#

It is well known that all the unitary operators 
in the same Hilbert space form a unitary group.  
Similarly, we have the following
group theorem for the  RUT:

$ Theorem~II$~~~All the operators satisfying Eq. (2) 
in the same Hilbert space 
form a semigroup.

$Proof$: ~In this proof, the main task is to prove that
for two arbitrary operators $U_{1},~ U_{2} \in$ RUT 
in the same Hilbert space, that $U=U_{1} U_{2}$ should also belong 
to RUT. It is evident that $UU^{\dagger}=1$. We now prove
$U^{\dagger}U\neq1$. Let
$W_{1}=U_{1}^{\dagger}U_{1},~W_{2}=U_{2}^{\dagger}U_{2}$. Then
$W\equiv U^{\dagger}U=U_{2}^{\dagger}W_{1}U_{2}$. The 
Corollary I. a tells us that for $U_{2}$, there exists at least
one ket vector 
$|\psi\rangle\neq 0$ to satisfy the equation $U_{2}|\psi\rangle=0$. 
Which means $W|\psi\rangle=0$, and $W\neq1$. Therefore,
$U\in$ RUT. The associative law is easily proved. 
Since all the RUT have not strict inverses, we conclude 
that all these RUT in the same Hilbert space 
form a semigroup, which is denoted as RUSG.
~\#

The above two theorems reveal some common properties of RUT, 
which are independent of the concrete structure of RUT.
In order to study the structure of RUT in detail, 
we now introduce a useful quantity: the $Kernel ~(K)$ of 
the operator $U\in$ RUT: 
\begin{equation} 
K=\{|\Phi_{i}^{0}\rangle,~i=1,2,\cdots\},~~ \mbox{for any}~
 |\Phi_{i}^{0}\rangle\in K,~ ~U|\Phi_{i}^{0}\rangle=0.
\end{equation}
Evidently, the subset $\{|\Psi^{0}\rangle\}$  in  Theorem I and the
arbitrary linear combinations of those elements in  
$\{|\Psi^{0}\rangle\}$ 
all belong to $K$. Therefore, $K$ has infinite number of elements. 
Here we introduce  another quantity--$Basic ~Kernel$,
which includes only the base ket vectors of $K$, namely, these ket 
vectors are orthogonal with each other.
In most cases, the  Basic Kernel has finite number of elements.
We sometimes simply call the Basic Kernel as Kernel.
For convenience, we usually denote
$K= \{|\Phi^{0}\rangle\}$, or   $K= |\Phi^{0}\rangle$ 
in the following presentation.
  
The Kernel distinguishes clearly the RUT from the unitary 
transformation,
since for any unitary transformation,  $K$ is an empty set.
In a certain sense,  $K$ can be taken as a measure of the non-
unitarity
of the operator $U\in$ RUT: the fewer  elements  $K$ has,
the more nearly unitary  $U$ is. The latter discussion will show that
 $K$ directly determines the applicability of RUT
to a physical system. 

Since RUT has been associated with 
a structure of semigroup, we now prove the following
Theorem about the Kernel  of
 several RUT's product: 

$ Theorem~III$ ~~~For those $U_{1},~U_{2},\cdots,U_{n},~\cdots\in$ 
RUSG,
the Basic Kernel of  operator 
\begin{equation}
U(1,2,\cdots,n)= 
U_{1}U_{2}\cdots U_{n}
\end{equation} is exactly
\begin{equation}
K=\left\{|\Phi^{0}(n)\rangle, 
~U_{n}^{\dagger}|\Phi^{0}(n-1)\rangle,
~U_{n}^{\dagger}U_{n-1}^{\dagger}|\Phi^{0}(n-2)\rangle, ~
U_{n}^{\dagger}\cdots 
U_{3}^{\dagger}~U_{2}^{\dagger}|\Phi^{0}(1)\rangle
\right\},
\end{equation}
where $|\Phi^{0}(m)\rangle$ is the Basic Kernel of $U_{m}$. 
(The bracket is omitted here)

$Proof$:~ The proof includes three steps.
First, let
\begin{equation}
W\equiv U^{\dagger}(1,2,\cdots,n) ~U(1,2,\cdots,n)=
U_{n}^{\dagger}\cdots U_{3}^{\dagger}~U_{2}^{\dagger}
W_{1}U_{2}U_{3}\cdots U_{n},
\end{equation}
using  Theorem I, we obtain that 
\begin{equation}
W|\Phi^{0}(n)\rangle= 
W\left[ U_{n}^{\dagger}|\Phi^{0}(n-1)\rangle\right]=\cdots=
W\left[ U_{n}^{\dagger}\cdots U_{3}^{\dagger}~
U_{2}^{\dagger}|\Phi^{0}(1)\rangle\right]=0.
\end{equation}
Therefore, all the elements $|\Phi^{0}(n)\rangle, 
~U_{n}^{\dagger}|\Phi^{0}(n-1)\rangle,\cdots, ~
U_{n}^{\dagger}\cdots U_{3}^{\dagger}~U_{2}^{\dagger}
|\Phi^{0}(1)\rangle$ in Eq. (12) belong to
the Kernel of $U(1,2,\cdots,n)$.

The second step, it is easily proved that
all the elements in Eq. (12) are orthogonal with each other.
We omit the proof here.

The third step, we will show that an arbitrary element
belonging to the Kernel of $U$ is uniquely
determined by the elements in Eq. (12). Supposing an arbitrary 
vector $ |\alpha\rangle$ that satisfies 
$W|\alpha\rangle=0$, by Eq. (13), we have
\begin{equation}
W_{1}U_{2}U_{3}\cdots U_{n}|\alpha\rangle=0.
\end{equation}
We conclude from this equation that there are
only two possibale $ |\alpha\rangle$ as follows : 

(i)~ $U_{2}U_{3}\cdots U_{n}|\alpha\rangle\in|\Phi^{0}(1)\rangle$. 
From  Theorem I, we obtain that
$U_{3}\cdots U_{n}|\alpha\rangle=c_{1}U_{2}^{\dagger}
|\Phi^{0}(1)\rangle+c_{2}|\Phi^{0}(2)\rangle$, 
where $c_{i}$ are the parameters commutative with
all  $U_{i}$. By this result, we obtain further  
$U_{4}\cdots 
U_{n}|\alpha\rangle=c_{1}U_{3}^{\dagger}U_{2}^{\dagger}
|\Phi^{0}(1)\rangle+
c_{2}U_{3}^{\dagger}|\Phi^{0}(2)\rangle+c_{3}|\Phi^{0}(3)\rangle$. 
Following the same analysis, we eventually get to that 
$|\alpha\rangle=c_{1}U_{n}^{\dagger}\cdots U_{3}^{\dagger}
~U_{2}^{\dagger}|\Phi^{0}(1)\rangle+
c_{2}U_{n}^{\dagger}\cdots U_{3}^{\dagger}|\Phi^{0}(2)\rangle+\cdots
c_{n}|\Phi^{0}(n)\rangle$, which evidently belongs to $K$, where 
$\sum\limits_{i=1}^{n}|c_{i}|^{2}=1$.

(ii)~ The other possibility is that $U_{l}\cdots U_{n}|\alpha\rangle=0$,  
while $U_{l+1}\cdots 
U_{n}|\alpha\rangle=d_{l}|\Phi^{0}(l)\rangle\neq 0$, 
where $2\leq l$.
Using the same discussion as in (i), we obtain that $|\alpha\rangle=
d_{l}U_{n}^{\dagger}\cdots U_{l+1}^{\dagger}|\Phi^{0}(l)\rangle+
\cdots+
d_{n}|\Phi^{0}(n)\rangle$, which still belongs to $K$
in Eq. (12). 

The above analysis leads to a conclusion that the 
expression (12) uniquely determines the Kernel of $U(1,2,\cdots,n)$. 
~\#

With the help 
of above several theorems, we now attempt to 
establish a connection between RUT 
and  $\Lambda$ in Eq. (1). It seems that the above properties 
of RUT are independent of $\Lambda$,
even though RUT is initiated by its 
application to  $\Lambda$.  The following Theorem IV 
expresses a sufficient condition, under which we can apply RUT to
 $\Lambda$: 

$ Theorem~IV$ ~~~ If the  Kernel of $U$,  
$ K=\{|\Phi_{i}\rangle\}$, is linearly isomorphic with the  set
$\{|\psi_{l}\rangle,~~l=1,2, \cdots,s\}$, which
is a subset of the total eigenkets $\{|\psi_{n}\rangle,
~~n=1,2,
\cdots,\infty\}$ of the operator $\Lambda$ in Eq. (1), 
then $U$ can be taken as a unitary transformation to the 
remaining subspace of $\Lambda$ without changing the spectrum.

$Proof$:~ We complete the proof by two steps. First, 
from the proposition we know that
$|\psi_{l}\rangle=\sum\limits_{i=1}^{s}d_{li}|\Phi_{i}\rangle$, 
~$l\leq s$. By which we obtain further
\begin{equation}
\left\{
\begin{array}{l}
U|\psi_{l}\rangle=0,~~l=1,2, \cdots,s,\\[.05in]
U|\psi_{m}\rangle\neq 0,~~m=s+1, \cdots, \infty.
\end{array}
\right.
\end{equation}
Above equations indicate that the total state space of  
$\Lambda$ can be divided into two parts, which are simply denoted 
as
$\{|\psi_{l}\rangle\}$ and $\{|\psi_{m}\rangle\}$, respectively.
The subspace $\{|\psi_{l}\rangle\}$ can  be determined by 
both of the Kernel of $U$ and $\Lambda$,
but the remaining subspace  $\{|\psi_{m}\rangle\}$ 
is independent of $K$.

Second, we now deal with  the remaining subspace 
$\{|\psi_{m}\rangle\}$. Let
$|\psi_{m}'\rangle=U^{\dagger}U |\psi_{m}\rangle$. 
It follows from Eq. (2)
that $U (|\psi_{m}'\rangle-|\psi_{m}\rangle)=0$. 
Equation (14) implies  
\begin{equation}
|\psi_{m}'\rangle=|\psi_{m}\rangle+\sum\limits_{l=1}^{s}
c_{ml}|\psi_{l}\rangle,
\end{equation}
where $c_{ml}$ are parameters to be determined. 
$|\psi_{m}'\rangle$ is further left-multiplied by a bra 
$\langle\psi_{l'}|$, $l'\leq s$:
\begin{equation}
\langle\psi_{l'}|U^{\dagger}U|\psi_{m}\rangle=
\langle\psi_{l'}|\psi_{m}\rangle+
\sum\limits_{l=1}^{s}c_{ml}\langle\psi_{l'}|\psi_{l}\rangle,
\end{equation}
from  Eq. (16) and the orthogonality of the ket vectors 
$\{|\psi_{n}\rangle\}$ 
indicated in Eq. (1), 
we obtain  $c_{ml}=0$ in Eq. (17).
Then, $|\psi_{m}'\rangle=U^{\dagger}U |\psi_{m}\rangle
=|\psi_{m}\rangle$. 
This result evidently shows that in the subspace
$\{|\psi_{m}\rangle\}$, $U$ acts as a unitary operator.
 
It is easy to  prove that
$\Lambda'=U\Lambda
U^{\dagger}$ has the same spectrum as $\Lambda$
in the  subspace $\{|\psi_{m}\rangle\}$.  
The new eigenkets of $\Lambda'$, $|\Psi_{m}\rangle= 
U|\psi_{m}\rangle$, 
are also complete, since 
\begin{equation}
\sum\limits_{m=s+1}^{\infty}
|\Psi_{m}\rangle\langle\Psi_{m}|
=U\{1-\sum\limits_{l=1}^{s}|\psi_{l}\rangle\langle\psi_{l}|
\}U^{\dagger}=UU^{\dagger}=1.
\end{equation}
Moreover, by that $U^{\dagger}|\Psi_{m}\rangle=|\psi_{m}\rangle$, 
we obtain the eigenkets of $\Lambda$. ~\#

The proposition in  above  Theorem IV  that
the Kernel of $U$ should be isomorphic with a subspace of  
$\Lambda$ is very strong. How to weaken  this proposition 
is still under investigation. 

Notice that in  the proof of  Theorem IV,
we have employed the eigenstates of $\Lambda$.
Therefore, one may wonder how to use the
RUT method to obtain the spectrum of $\Lambda$, 
provided  we do not know these eigenstates at first.
To answer this question, 
we should keep in mind that even when  
using the unitary transformation to solve a problem, there still
is not a widely accepted rule on how to construct
a unitary transformation, except some well-known problems  
in Ref. [2].
However, for the RUT method, based on  Theorem IV,
we conclude a general way on how to apply RUT to a concrete 
problem, 
there are several steps described as follows: 

Consider a Hermitian operator such as $\Lambda$ in Eq. (1). 
To solve the eigenvalue equation, 
we first attempt to construct a transformation $U$ 
in the same Hilbert space as $\Lambda$ to simplify it.
For the purpose, $U$ is constructed to be a
right-unitary operator, its Kernel is easily obtained. 
Then we can directly check whether all the elements 
in the Kernel (or the combinations of the
those elements in Kernel) are the eigenstates of $\Lambda$. 
If they are, the Kernel is  called 
$proper~ Kernel$, which  guarantees that we can
directly follow  Theorem IV to
take $U$ as a unitary transformation to  
transform  $\Lambda$ into $\Lambda'=U\Lambda U^{\dagger}$,
without changing the spectrum.
 
We should notice that in the above steps of using RUT,
we need not know the eigenkets of $\Lambda$ at first. The most 
important step is to construct a RUT with a $proper~ Kernel$,
as  Theorem IV required. 
In practice, the number of the basic elements 
in Kernel should be as few as possible, because: 
i) with a smaller Kernel, it is easier for us to check 
whether it is a $proper$ $Kernel$; 
ii) From  Theorem I, we know that in comparison, 
the smaller the Kernel, the more nearly unitary  the RUT is. 

In Section {\bf III}, we will follow the above steps to treat 
various  matter-radiation interactions.\\

\begin{center}
\subsection*{2.2.  Non-stationary case}
\end{center}

We now consider such a case that 
 the Hamiltonian of a system represented by
$\Lambda$ varies with time. The evolution of the 
state obeys the 
Schr$\ddot{o}$dinger equation  $(\hbar=1)$
\begin{equation}
\Lambda(t)|\psi(t)\rangle=i\frac{\partial}{\partial t}
|\psi(t)\rangle,~~~
|\psi(0)\rangle=|\psi_{0}\rangle.
\end{equation}
We further assume  that there exists another 
variable $\hat{R}$ in $\Lambda(t)$: 
$\Lambda(t)\equiv\Lambda(\hat{R},t)$,
and the state space of $\hat{R}$ is an infinite dimensional 
Hilbert space. 
Consider an operator $U\in$ RUT in the same Hilbert space 
as $\hat{R}$, and independent of time.
$U$ has a finite  Kernel: $K=\{|\Psi_{i}^{0}\rangle~,
 i=1,2,\cdots,s\}$, and a unitary subspace
$\{|\Psi_{i}^{1}\rangle~,
 i=s+1,s+2,\cdots,\infty\}$ as indicated in  Theorem I.
In order to apply $U$ to the system $\Lambda(t)$, we
present the following Theorem which is essentially
a revision of Theorem IV:

$Theorem ~V: $~~~For $U\in$ RUT, if all the elements or the linear
combinations
of those elements in the Kernel of $U$ are exactly the eigenstates
of $\Lambda(t)$ in Eq. (20), i.e.,
\begin{equation}
\Lambda(t)|\xi_{n}\rangle=\lambda_{n}(t)|\xi_{n}\rangle, 
~~m=1,2,\cdots, s
\end{equation}
where $|\xi_{n}\rangle=\sum\limits_{m=1}^{s}d_{nm}
|\Psi_{m}^{0}\rangle$, 
the matrix $(d_{nm})$ has an inverse, and 
$t$ in Eq. (21) is taken as a constant parameter, 
then  $U$  can be applied 
to the remaining subspace of $\Lambda(t)$ without changing the 
Schr$\ddot{o}$dinger equation.

$Proof$:~~Since the matrix $(d_{nm})$ 
has an inverse,  two subspaces
$\{|\xi_{n}\rangle, ~n=1,2,\cdots,s\}$ and 
$\{|\Psi^{0}_{n}\rangle, ~n=1,2,\cdots,s\}$ 
are isomorphic with each other.
 Theorem I indicates that there is a complete set of ket vectors 
$\{|\xi_{n}\rangle, |\Psi_{i}^{1}\rangle,
~n=1,2,\cdots,s;~ i=s+1,s+2,\cdots,\infty\}$ 
for $W\equiv U^{\dagger} U$, where the ket vectors
are assumed to be orthogonal with each other. Then the wave 
function
of $\Lambda(t)$ can be expressed as 
\begin{equation}
|\psi(t)\rangle=\sum\limits_{m=1}^{s}
f_{m}(t)|\xi_{m}\rangle+\sum\limits_{n=s+1}^{\infty}
g_{n}(t)|\Psi_{n}^{1}\rangle,
\end{equation}
where the time-dependent parameters
$f_{m}(t) $ and $g_{n}(t)$ are to be determined.
Taking this expression (22) into Eq. (20), 
by Eq. (21) and the orthogonality
of those ket vectors, we obtain that the parameters
$f_{m}(t) $ satisfy the  equation 

\begin{equation}
\frac{\partial}{\partial t}f_{m}(t)=-i\lambda_{m}(t)f_{m}(t),
~~m=1,2,\cdots,s,
\end{equation}
which is solved as
\begin{equation}
f_{m}(t)=f_{m}(0)~\exp\left[-
i\int\limits^{t}_{0}\lambda_{m}(t')dt'\right],
~~m=1,2,\cdots,s.
\end{equation}
In the remaining subspace $\{ |\Psi_{i}^{1}\rangle,~ 
i=s+1,s+2,\cdots,\infty\}$, since $U$ acts as a unitary transformation,
the Schr$\ddot{o}$dinger equation (20) can be turned into
\begin{equation}
\Lambda'(t)|\psi(t)\rangle^{\prime}=i\frac{\partial}
{\partial t}|\psi(t)\rangle^{\prime},
\end{equation}
where $\Lambda'(t)=U\Lambda(t)U^{\dagger}$, and 
$|\psi(t)\rangle^{\prime}=\sum\limits_{l=s+1}^{\infty}
g_{l}(t)U|\Psi_{l}^{1}\rangle$. In the transformed frame 
$\Lambda'(t)$,
the parameters $g_{n}(t) $ are obtained to satisfy the equation
\begin{equation}
\frac{\partial}{\partial t}g_{n}(t)=-i\sum\limits_{l=s+1}^{\infty}
g_{l}(t)\langle\Psi_{n}^{1 \prime}
|\Lambda^{\prime}(t)|\Psi_{l}^{1}\rangle^{\prime},
~~n=s+1,s+2,\cdots,\infty,
\end{equation}
where $|\Psi_{l}^{1}\rangle^{\prime}=U|\Psi_{l}^{1}\rangle$.
The equations (25) and (26) evidently show that $U$  can be applied 
to the remaining subspace of $\Lambda(t)$. ~\#

The condition Eq. (21) is the key point of this theorem, by which 
the total state space of $\Lambda(t)$ can be divided into two parts:
one part is determined by the Kernel of $U$; the other part 
can be treated by $U$ as a unitary transformation. In the above 
proof,  
the subspace  $\{ |\Psi_{i}^{1}\rangle,~ i=s+1,s+2,\cdots,\infty\}$ 
has been used to expand the wave function. In fact, for a concrete 
problem,
we do not know this subspace at first. Instead, we can
directly apply $U$ to $\Lambda(t)$ as shown
in Eq. (25) to obtain this subspace,
provided the condition Eq. (21) is satisfied already.

The above $U\in$ RUT is assumed to be independent of time, 
since $\hat{R}$ is an time-independent variable in $\Lambda(t)$. 
One can extend  Theorem V into the case of many variables  without
difficulty.\\

\begin{center}
\subsection*{2.3. Inverses of bosonic operators and examples of RUT}
\end{center}

As one of the infinite dimensional Hilbert space, the Fock space has 
been
widely used in quantum physics. In order to construct 
RUT conveniently in the Fock space,
we employ the recently introduced inverses of boson creation and 
annihilation 
operators $a^{\dagger} $ and $a$, $[a,~a^{\dagger}]=1$. 
Mehta $et~al.$ \cite{arvind} found that  
$a^{\dagger}$ has a left inverse $b$, and  
$a$ has a right inverse $b^{\dagger}$,
\begin{equation}
b a^{\dagger}=a b^{\dagger}=1. 
\end{equation}
The properties of $b,b^{\dagger}$ and their applications to squeezed 
states,
and the  M$\ddot{o}$bius transformation have 
obtained detailed studies in Ref. [12].
In Ref. [9], we showed that $b$ and $b^{\dagger}$ can be formally
expressed in terms of $a$ and $ a^{\dagger}$,
\begin{equation}
b=(\frac{1}{a a^{\dagger}}) a, ~~~~
b^{\dagger}=a^{\dagger} (\frac{1}{a a^{\dagger}}).
\end{equation}
From Eq. (28), we obtain further:  $bb^{\dagger}=1/a a^{\dagger}$.
 Using this relation, we  
arrive  an interesting result:  
$a$ and $ a^{\dagger}$ can be formally 
expressed in terms of  $b$ and $b^{\dagger}$ too,
\begin{equation}
a=(\frac{1}{b b^{\dagger}}) b, ~~~~
a^{\dagger}=b^{\dagger} (\frac{1}{b b^{\dagger}}).
\end{equation}
We know  in the scheme of second quantization, an arbitrary 
operator can be expanded  by
$a$ and $a^{\dagger}$, namely,
$a$ and $a^{\dagger}$ form a $complete~ operator ~set$. 
Equation (29)  implies that any operator can also 
be expanded by $b$ and $
b^{\dagger}$. Therefore, 
$b$ and $b^{\dagger}$   form a
$complete~ operator ~set 
$~ too, which is connected to that of $a$ and $a^{\dagger}$ by 
the nonlinear transformations
(28) or (29). 

Equation (28) significantly simplifies the calculation
in the representation of  $b$ and $b^{\dagger}$.  
Various results in Ref. [12] such  as the analytic studies 
of $b$ and $b^{\dagger}$ in the representation of Bargmann 
space are easily obtained  by using Eqs. (28) and (29).
For  example,  using Eq. (28),
one can prove
the following useful formulas:
\begin{eqnarray}
b^{k}b^{\dagger k}=\frac{1}{a^{k}a^{\dagger k}}
=\frac{1}{(N+1)(N+2)\cdots(N+k)}\equiv\frac{N!}{(N+k)!},\\[.15in]
b^{\dagger k}a^{k}=\sum\limits_{n=0}^{\infty}b^{\dagger k}|n\rangle
\langle n|a^{k}=\sum\limits_{n=0}^{\infty}|n+k\rangle
\langle n+k|=1-\sum\limits_{n=0}^{k-1}|n\rangle
\langle n|,
\end{eqnarray}
where $N=a^{\dagger} a$. 

We now look at the well-known operator---Phase operator (phasor)
$F_{1}$ as shown in Eq. (3),
which is initiated from the quantization of phase factor 
in quantum mechanics \cite{phasor}. In this paper we will use only 
the 
photon-lowering property of these phase operators,
and that the phase properties of the electromagnetic 
field are not calculated.
The higher order phase operators are defined as:
\begin{equation}
F_{k}\equiv (F_{1})^{k}=\frac{1}{\sqrt{a^{k}a^{\dagger k} }}a^{k},
~~~F_{k}^{\dagger}\equiv (F_{1}^{\dagger})^{k}
=a^{\dagger k}\frac{1}{\sqrt{a^{k}a^{\dagger k}}}.
\end{equation}
In the representation of $b$ and $b^{\dagger}$, 
$F_{1}$ and $F_{1}^{\dagger}$ are found to have the same forms
as Eq. (3),
\begin{equation}
F_{1}=\frac{1}{\sqrt{bb^{\dagger}}}b,~~~
F_{1}^{\dagger}=b^{\dagger}\frac{1}{\sqrt{bb^{\dagger}}}.
\end{equation}  
By Eq. (30), we have 
\begin{equation}
F_{k}=\frac{1}{\sqrt{b^{k}b^{\dagger k} }}b^{k},
~~~F_{k}^{\dagger}=b^{\dagger k}\frac{1}{\sqrt{b^{k}b^{\dagger k}}}.
\end{equation} 
 
With these preparations, we now prove the following theorem:

$ Theorem~VI$: ~~~All the phase operators defined by
Eqs. (3) and (32) form a subgroup of the  
right-unitary semigroup (RUSG): $
\{1, ~F_{k}, ~k=1,2,\cdots\}\subset$ RUSG.

$Proof$:~~ It is clear that $ F_{k}F_{k}^{\dagger}=I$. 
On the other hand, using Eq. (31), we obtain 
\begin{equation}
F_{k}^{\dagger}~ F_{k}=
I-\sum\limits_{n=0}^{k-1}|n\rangle\langle n|\neq I.
\end{equation}
These mean that $ F_{k}\in$ RUT. Moreover, 
$ F_{k} F_{l}= F_{k+l}\in$ RUT. 
 We therefore conclude that
 $\{F_{k},~~k=1,2,\cdot, \infty\}$ form an Abelian subgroup of RUSG.
The Kernel of the element $F_{k}$ is directly obtained  as $
K(F_{k})=\{|0\rangle,~|1\rangle, \cdots~|k-1\rangle\}$.~\#

Using this explicit example of RUT, one can easily check  
the theorems presented above. We are not going to dig deeper the 
theory
of RUT here. It is important to see how to apply RUT to the physical
system.\\

\begin{center}
\section*{III. Application of RUT to the system of many-level
atoms interacting with quantized radiation fields}
\end{center}

Under the rotating-wave approximation, the interaction of two-level
atoms with the quantized radiation field
is described by the Jaynes-Cummings (JC) model \cite{jcm},
which has been extensively applied in quantum optics, quantum 
electronics, etc. Various modifications and
generalizations to the original JC model have been made to
approach quantum effects such as quantum collapses and 
revivals of atomic coherence \cite{collaps}, squeezing phenomenon
\cite{squeez} and so on.  These
JC models have  a common attractive property that they all can be 
solved 
exactly. Since the supersymmetric structure 
was found to exist the JC model,
the JC model is viewed as a generalization of the supersymmetric 
harmonic oscillator system, and its solvability may be 
interpreted in terms of supersymmetric breaking \cite{haymaker}. 
Using a deformed oscillator algebra, 
Bonatsos $et~ al$. \cite{bona} gave a unified solvable 
formulation of various JC Hamiltonians.  
Yu $ et~ al$. \cite{yu}  pointed out further that there embeds 
an unusual $su(2)$ algebraic structure in these JC Hamiltonians.  

The method mentioned above are valid to the two-level stationary
JC models only. It is difficult to extend these methods to 
the time-dependent JC models, or to the case where the atoms 
have higher ($>2$) energy levels. 
In a previous paper \cite{tang}, we employed a
right-unitary transformation (RUT) to solve exactly a generalized 
two-level JC model and pointed out that there exists
a geometric phase in the model.  

In this section, we will show that the RUT method 
developed  above  can be generally
applied to various forms of atom-radiation interactions, where the 
atoms
can have two, three and higher energy levels in different 
configurations,
and the radiation field can be one-mode or two-mode cavity. 
This section includes four subsections. 
In Subsection {\bf 3.1}, by  constructing a RUT, we  follow  Theorem 
V
to solve such a model that the  two-level atoms interact with
a radiation field which is dependent on time.
This model is called nonstationary  JC model. 
The importance of three-level matter-radiation interactions
has been pointed out in the Introduction. However, 
some interesting characteristics  such as the supersymmetric
structure, and the  $su(2)$ structure are not 
embedded in the three-level models. 
There still is not a generally accepted method to treat
these models, according to our knowledge. 
In Subsections {\bf 3.2, 3.3} and {\bf 3.4}, following  
Theorem IV, we will show that various interactions of 
many-level atoms with one, or two-mode cavity 
can be unitedly treated by the RUT method, 
where the energy of atoms can be two-level and  three-level 
with $\Lambda, ~V$ and $\equiv$-configurations.
The procedures of the treatment are as 
simple as those in the two-level JC model.\\

\begin{center}
\subsection*{ 3.1.~ Two-level nonstationary Jaynes-Cummings 
model}
\end{center}

The approach of the stationary two-level JC models
by using the RUT method can be found in Ref. [9]. 
In this subsection,
we consider such a case that the radiation field interacting with the 
two-level atoms varies with the external source. 
Through this model, we hope 
to achieve the goal of controlling the effects such as atomic 
collapse and revival, 
and the statistics of photon number through the external
source.  
Under the rotating-wave approximation, the Hamiltonian of  the 
system with  density $\rho(N)$-dependent
multi-photon ($l$) interaction has the following general form 
($\hbar=1$):
\begin{equation}
H(t)=\omega a^{\dagger} a+ \frac{1}{2}\omega_{0}\sigma_{3}+   
a^{\dagger l}\rho(N)\gamma^{*}(t)\sigma_{-}+
\rho(N) a^{l}\gamma(t)\sigma_{+},
\end{equation}
where $\omega$ and $\omega_{0}$ are the field and atomic
transition frequencies, respectively. $\sigma_{\pm}
=(\sigma_{1}\pm i\sigma_{2})/2$, where $\sigma_{1},~ \sigma_{2}, 
~\sigma_{3} $ 
are three Pauli matrices. $\gamma(t)$ represents the change 
of radiation field with the external source, $\gamma(t)\neq0$.
The detuning
$\Delta=\omega_{0}-l\omega$ should satisfy the condition:
$|\Delta|\ll \omega_{0}, \omega$, in order to preserve the
reliability of the rotating-wave approximation. The evolution of 
the system with time is described by the Schr$\ddot{o}$dinger 
equation
\begin{equation}
H(t)|\Psi(t)\rangle=i\frac{\partial}{\partial t}|\Psi(t)\rangle,~~~
|\Psi(0)\rangle=|\Psi_{0}\rangle.
\end{equation}
To solve this equation, we construct the following operator $U$
\begin{equation}
U=\left (
\begin{array}{cc}
1,& 0\\
0,& F_{l}
\end{array}
\right ),
\end{equation}
where $F_{l}=\frac{1}{\sqrt{b^{l}b^{\dagger l} }}b^{l}$ 
is the phase operator given by Eq. (34). 
It should be mentioned that this $U$ is expressed in the two-
dimensional
representation of the Pauli matrices, where the parameters are 
phase operators. From  Theorem VI, we know
that $U$  has the  properties 
\begin{equation}
\left\{
\begin{array}{l}
U U^{\dagger}=I,\\
U^{\dagger} U= \left (
\begin{array}{cc}
1,& 0\\[.10in]
0,& 1-\sum\limits_{k=0}^{l-1}|k\rangle\langle k|
\end{array}
\right),
\end{array}
\right.
\end{equation}
which indicate that $U$ belongs to RUT, and the Kernel of $U$ is 
\begin{equation} 
K=\left\{|\psi^{0}_{k}\rangle=
\left(
\begin{array}{c}
0\\
|k\rangle
\end{array}
\right),~~~k=0,1,\cdots l-1 \right\}.
\end{equation}
One can check that the vectors $|\psi^{0}_{k}\rangle, 
~k=0,1,\cdots l-1,$ in $K$  are
exactly the eigenstates of $H(t)$, where the eigenvalues 
\begin{equation}
E^{0}_{k}=\omega k-\omega_{0}/2.
\end{equation}
These results evidently show  that
$U$ is covered by  Theorem V.  We now follow  Theorem V to
solve the equation (37). The wave function
$|\Psi(t)\rangle$ can be divided into two parts:
\begin{equation}
|\Psi(t)\rangle=|\Psi^{0}(t)\rangle+|\Psi^{1}(t)\rangle,
\end{equation}
where $|\Psi^{0}(t)\rangle$ is related to the Kernel and 
given by Eq. (24) as
\begin{equation}
|\Psi^{0}(t)\rangle=\sum\limits_{k=0}^{l-1}f_{k}e^{-i(
\omega k-\omega_{0}/2)t}|\psi^{0}_{k}\rangle.
\end{equation}
the other part $|\Psi^{1}(t)\rangle$ is determined by the unitary
subspace of $U$. To obtain $|\Psi^{1}(t)\rangle$,
we make a transformation:  $H'(t)= UH(t)U^{\dagger}$, 
that is
\begin{equation}
H'(t)=\left (
\begin{array}{cc}
\omega N+\frac{1}{2}\omega_{0},& g_{l}(N)\gamma(t)\\[.15in]
g_{l}(N)\gamma^{*}(t), & \omega (N+l)-\frac{1}{2}\omega_{0}
\end{array}
\right ),
\end{equation}
where 
$g_{l}(N)=\rho(N) [(N+l)!/N!]^{\frac{1}{2}}$. 
The Hamiltonian now turns out to be a function of the 
photon number $N$, 
where the creation and annihilation of photon have been erased
by the transformation $U$. 
 
We know that in the original frame $H(t)$,
the transition from the atomic state $2$ to the state $1$
is induced by the dipole term $a^{\dagger l}\rho(N)\gamma^{*}(t)$,
and the transition is always achieved by  creating
$l$ photons. On the other hand, 
the transition from the state $1$ to
state $2$ is achieved by the annihilating the same  $l$ photons.
However, in the new frame $H'(t)$, 
all of these transitions are caused not
by the creating or annihilating  photon.
The transition induced 
by dipole can only happen between those states having the 
same number of photons, as shown in  Fig. 1. 
 Therefore, the states with different photon 
numbers can be treated independently.  

$H'(t)$  is further rearranged into the following compact form,
\begin{equation}
H'(t)=B_{0}-
\vec{\mu}\cdot\vec{B}(t),
\end{equation}
where $B_{0}=\omega (N+l/2),~ 
B_{1}(t)=\mbox{Re}[\gamma(t)]g_{l}(N),
~ B_{2}(t)=-\mbox{Im}[\gamma(t)]g_{l}(N), ~B_{3}=
(\omega_{0}-l)/2$, ~$ \vec{\mu}=-\vec{\sigma}.$~

The above result [Eq. (45)] reveals that in the transformed frame,
the Hamiltonian   $H'(t)$ of the nonstationary JC model 
describes  the motion of a spin-$\frac{1}{2}$ atom 
under an  effective magnetic field $\vec{B}(t)$, which 
is a function of the photon
number $N$ and time $t$. 

The evolution of subspace $|\Psi^{1}(t)\rangle^{\prime}$ 
is described by the Schr$\ddot{o}$dinger equation (23)  
\begin{equation}
H'(t)|\Psi^{1}(t)\rangle^{\prime}=
i\frac{\partial}{\partial t}|\Psi^{1}(t)\rangle^{\prime},
\end{equation}
where $|\Psi^{1}(t)\rangle^{\prime}=\sum\limits_{k=0}^{\infty}
\left(
\begin{array}{c}
\alpha_{k}(t)|k\rangle\\
\beta_{k}(t)|k\rangle
\end{array}
\right)$, the coefficients $\alpha_{k}(t)$ and $\beta_{k}(t)$
 satisfy the equations 
\begin{equation}
\left\{
\begin{array}{l}
\dot{\alpha_{k}}(t)=-i\alpha_{k}(t)(\omega k+\omega_{0}/2)-i
\beta_{k}(t)g_{l}(k)\gamma(t),\\[.15in]
\dot{\beta_{k}}(t)=-i\beta_{k}(t)[\omega (k+l)+\omega_{0}/2]-i
\alpha_{k}(t) g_{l}(k)\gamma^{*}(t),
\end{array}
\right.
\end{equation}\\
which are further arranged into
\begin{equation}\left\{
\begin{array}{l}
\ddot{\alpha_{k}}(t)+\dot{\alpha_{k}}(t) A'_{k}[l,~\gamma(t)]
+\alpha_{k}(t) B'_{k}[l,~\gamma(t)]=0,\\[.15in]
\ddot{\beta_{k}}(t)+\dot{\beta_{k}}(t) A'_{k}[l,~\gamma^{*}(t)]
+\beta_{k}(t) B'_{k}[l,~\gamma^{*}(t)]=0,
\end{array}
\right.
\end{equation}\\
where $A'_{k}[l,\gamma(t)]=i\omega(2k+l)-
\dot{\gamma}(t)/\gamma(t)$,~
$B'_{k}[l,\gamma(t)]=g_{l}^{2}(k)|\gamma(t)|^{2}-
i(\omega k+\omega_{0}/2)
\dot{\gamma}(t)/\gamma(t)-\omega^{2}(k^{2}+lk)-
\omega\omega_{0}/2+
\omega_{0}^{2}/4$.

For a given  $\gamma(t)$, by solving the above two equations,
one can obtain the evolution of the system with time. For instance,
we choose a special case,  
\begin{equation}
\gamma(t)=e^{i\lambda t},
\end{equation}
 to approach the dependence of some physical properties in 
this model on $\lambda$. With this particular $\gamma(t)$,
$\alpha_{k}(t)$ and $\beta_{k}(t)$ are solved to be
\begin{equation}\left\{
\begin{array}{l}
\alpha_{k}(t)=\alpha^{+}_{k}e^{i\omega_{k}^{+}t}
+\alpha^{-}_{k}e^{i\omega_{k}^{-}t},\\[.15in]
\beta_{k}(t)=-h_{k}\alpha^{+}_{k}
e^{i(\omega_{k}^{+}-\lambda)t}+
h^{-1}_{k}\alpha^{-}_{k}
e^{i(\omega_{k}^{-}-\lambda)t},
\end{array}
\right.
\end{equation}
where 
\begin{equation}
\begin{array}{l}
\omega_{k}^{\pm}=-\frac{1}{2}[\omega(2k+l)
-\lambda]\pm\sqrt{\frac{1}{4}(\omega_{0}-\omega l+\lambda)^{2}
+[g_{l}(k)]^{2}},\\[.15in]
h_{k}=
(\omega k+\frac{\omega_{0}}{2}+\omega_{k}^{+})/g_{l}(k),
\end{array}
\end{equation}\\
and the coefficients $\alpha^{\pm}_{k}$ 
and $\beta^{\pm}_{k}$ 
are determined by the initial state 
in the transformed frame:
\begin{equation}
|\Psi^{1}(0)\rangle^{\prime}=\sum\limits_{k=0}^{\infty}
\left(
\begin{array}{c}
(\alpha^{+}_{k}+\alpha^{-}_{k})|k\rangle\\[.10in]
(-h_{k}\alpha^{+}_{k}+h^{-1}_{k}\alpha^{-}_{k})|k\rangle
\end{array}
\right).
\end{equation} 
If the initial state in the transformed frame is chosen as
\begin{equation}
|\Psi^{1}(0)\rangle^{\prime}=\sum\limits_{k=0}^{\infty}
\left(
\begin{array}{c}
\mu_{k}|k\rangle\\[.10in]
\nu_{k}|k\rangle
\end{array}
\right),
\end{equation}
then the coefficients $\alpha^{\pm}_{k}$ and $\mu_{k},~ \nu_{k}$
 are related with each other by the equations
\begin{equation}\left\{
\begin{array}{l}
\alpha_{k}^{+}=\frac{1}{1+h_{k}^{2}}(\mu_{k}-h_{k}\nu_{k}),\\[.15in]
\alpha_{k}^{-}=\frac{h_{k}}{1+h_{k}^{2}}(h_{k}\mu_{k}+\nu_{k}).
\end{array}
\right.
\end{equation}
For convenience, we still use the initial state
as Eq. (52) in  the later calculations. 

Since the initial state is arbitrarily chosen, 
the equation (50) is in fact the general solution
of the non-stationary JC model in the transformed frame. From
 Theorem V, we know that the
solution can be directly turned into the $untransformed$ frame
by 
\begin{equation}
|\Psi^{1}(t)\rangle=U^{\dagger}|\Psi^{1}(t)\rangle^{\prime}.
\end{equation}
Combining this solution and the Kernel part in Eq. (43),
we obtain the exact solution of the non-stationary JC model 
for an arbitrary initial state
as\\
\begin{equation}
|\Psi(t)\rangle=\sum\limits_{n=0}^{l-1}f_{n}e^{-i(
\omega n-\omega_{0}/2)t}\left(
\begin{array}{c}
0\\
|n\rangle
\end{array}
\right)+
\sum\limits_{k=0}^{\infty}
\left(
\begin{array}{c}
\left[\alpha^{+}_{k}e^{i\omega_{k}^{+}t}
+\alpha^{-}_{k}e^{i\omega_{k}^{-}t}\right]|k\rangle,\\[.15in]
\left[-h_{k}\alpha^{+}_{k}e^{i\omega_{k}^{+}t}
+h^{-1}_{k}\alpha^{-}_{k}e^{i\omega_{k}^{-}t}\right]
e^{-i\lambda t}|k+l\rangle,
\end{array}
\right).
\end{equation}\\
Using the normalization condition:
$\langle\Psi(t)|\Psi(t)\rangle=1$, we know that all the coefficients 
should satisfy the equation
\begin{equation}
\sum\limits_{n=0}^{l-1}|f_{n}|^{2}+\sum\limits_{k=0}^{\infty}
\left[(1+h_{k}^{2})|\alpha^{+}_{k}|^{2}+
(1+h_{k}^{-2})|\alpha^{-}_{k}|^{2}\right]=1.
\end{equation}

The above solution (56) shows that the Kernel term is
independent of $\gamma(t)$, thereby a trivial term.
Our interest is in the second term in the solution (56), because
this term is not only related 
to $\gamma(t)$, but also determined by the unitary subspace
of $U$. We now  evaluate the following
physical quantities associated with
the non-stationary JC model by using RUT.\\

i) Statistics of the photon number

Let $\bar{n}^{\pm}(t)=\langle\Psi(t)|
(\sigma_{0}\pm\sigma_{3})N/2|\Psi(t)\rangle$, where $\sigma_{0}$
is a $2\times 2$ unit matrix, $\sigma_{3}$ is the third Pauli matrix.
For $\bar{n}^{+}$, since $U\left[(\sigma_{0}+\sigma_{3})N/2\right]
U^{\dagger}=(\sigma_{0}+\sigma_{3})N/2$,
we can treat it by RUT in the transformed frame,
\begin{equation}
\bar{n}^{+}(t)=\left[\langle\Psi^{1}(t)|U^{\dagger}\right]
(\sigma_{0}+\sigma_{3})N/2
\left[U|\Psi^{1}(t)\rangle\right]=\sum\limits_{k=0}^{\infty}
k\left[|\alpha^{+}_{k}|^{2}+
|\alpha^{-}_{k}|^{2}+\xi_{k}(t) \right].
\end{equation}
where 
\begin{equation}
\xi_{k}(t)=\alpha_{k}^{+*}\alpha_{k}^{-}
e^{-i(\omega_{k}^{+}-\omega_{k}^{-})t}
+\alpha_{k}^{+}\alpha_{k}^{-*}e^{i(\omega_{k}^{+}-\omega_{k}^{-})t}.
\end{equation}
By the same way, we have 
\begin{equation}
\bar{n}^{-}(t)=\sum\limits_{n=0}^{l-1}n|f_{n}|^{2}+
\sum\limits_{k=0}^{\infty}
(k+l)\left[|\alpha^{+}_{k}|^{2}h_{k}^{2}+
|\alpha^{-}_{k}|^{2}h_{k}^{-2}-\xi_{k}(t) \right].
\end{equation}
It is proper to refer to $\bar{n}^{+}(t)$ and $\bar{n}^{-}(t)$ 
as the mean photon number of the field for the cases
where the atom is found in the excited and ground
states, respectively. $\xi_{k}(t)$ is an important 
quantity in this model, it measures the collapses 
and revivals of the atomic inversion 
in a single mode with the frequency 
\begin{equation}
\Delta_{k}\equiv
\omega_{k}^{+}-\omega_{k}^{-}=
\sqrt{(\omega_{0}-\omega l+\lambda)^{2}
+4[g_{l}(k)]^{2}},
\end{equation}
$\Delta_{k}$ is usually called Rabi frequency,
where the parameter $\lambda$
obviously affects the Rabi frequency.\\
  
ii) The atomic inversion

The atomic inversion in the JC model is illustrated  by
the quantity $\bar{\sigma}_{3}=\langle\Psi(t)|\sigma_{3}
|\Psi(t)\rangle$. If we take the two-level atoms as the neutro 
particles
with spin-$\frac{1}{2}$, then $\bar{\sigma}_{3}/2$ measures
the mean spin value which varies with time. Since $\sigma_{3}$
is invariant under $U$, namely,
$U\sigma_{3} U^{\dagger}=\sigma_{3}$, we can directly calculate this 
quantity in the transformed frame. 
Using the normalization condition Eq. (57), we obtain
\begin{equation}
\bar{\sigma}_{3}=2 \sum\limits_{k=0}^{\infty}
\left[|\alpha^{+}_{k}|^{2}+|\alpha^{-}_{k}|^{2}+\xi_{k}(t)\right]-1.
\end{equation}
This result shows  that the atomic inversion is measured
by the same $\xi_{k}(t)$ as that appears in the statistics of photon 
number.
In order to evaluate explicitly the effect of 
$\lambda$ in the nonstationary JC model,
we choose the following special initial state without the Kernel part
for an example:
\begin{equation}
|\Psi(0)\rangle=\sum\limits_{k=0}^{\infty}e^{-|z|^{2}/2}
\left(
\begin{array}{c}
\frac{z^{k}}{\sqrt{2 k!}}|k\rangle\\[.10in]
\frac{z^{k}}{\sqrt{2 k!}}|k+l\rangle
\end{array}
\right).
\end{equation}
For this initial state, we have (see Appendix)
\begin{equation}
\xi_{k}(t)=-\left(\frac{x^{k}e^{-x}}{k!}\right)
\frac{(\Delta+\lambda)g_{l}(k)}{(\Delta+\lambda)^{2}+
4g^{2}_{l}(k)}\cos\left[\sqrt{(\Delta+\lambda)^{2}+
4g^{2}_{l}(k)}~t\right],
\end{equation}
\begin{equation}
|\alpha^{+}_{k}|^{2}+|\alpha^{-}_{k}|^{2}=\frac{x^{k}e^{-x}}{k!}
\left[\frac{1}{2}+\frac{(\Delta+\lambda)g_{l}(k)}{(\Delta+\lambda)^{2
}+
4g^{2}_{l}(k)}\right],
\end{equation}
where the detuning $\Delta=\omega_{0}-\omega l$, $x=|z|^{2}$.
With these results, $\bar{\sigma}_{3}$ is obtained to be
\begin{equation}
\bar{\sigma}_{3}=\sum\limits_{k=0}^{\infty}
\left(\frac{x^{k}e^{-x}}{k!}\right)
\frac{4(\Delta+\lambda)g_{l}(k)}{(\Delta+\lambda)^{2}+
4g^{2}_{l}(k)}\sin^{2}\left[\frac{1}{2}\sqrt{(\Delta+\lambda)^{2}+
4g^{2}_{l}(k)}~t\right].
\end{equation}
Equation (66) expresses exactly the collapse and revival 
of the atomic coherence in this nonstationary model,
where $\lambda$ plays the same role as the detuning $\Delta$. 
By changing $\lambda$, we can effectively control the
atomic inversion of the system, thus meet our original
purpose of studying the properties of the atomic system 
via the external source.
If it is designed: $\lambda\approx -\Delta$,
the atomic inversion disappears for the initial state (63). 
On the other hand, even if $\Delta=0$, from  Eq. (66) we
know that the atomic inversion still exists for the external 
source $\lambda\neq 0$,
\begin{equation}
\bar{\sigma}_{3}= \sum\limits_{k=0}^{\infty}
\left(\frac{x^{k}e^{-x}}{k!}\right)
\frac{4\lambda g_{l}(k)}{\lambda^{2}+
4g^{2}_{l}(k)}\sin^{2}\left[\frac{1}{2}\sqrt{\lambda^{2}+
4g^{2}_{l}(k)}~t\right].
\end{equation}\\

iii) Energy

The mean energy of the system is
 $\bar{E}(t)=\langle\Psi(t)|H(t)|\Psi(t)\rangle$.
For the state (56), we divide $\bar{E}(t)$ into three terms,
\begin{equation}
\bar{E}(t)=\langle\Psi^{0}(t)|H(t)|\Psi^{0}(t)\rangle
+\left[\langle\Psi^{1}(t)|H(t)|\Psi^{0}(t)\rangle+h.c.\right]+
\langle\Psi^{1}(t)|H(t)|\Psi^{1}(t)\rangle,
\end{equation}
where the first term in $\bar{E}(t)$ is easily obtained to be 
\begin{equation}
\bar{E}^{0}(t)=\sum\limits_{n=0}^{l-1}(\omega n-
\omega_{0}/2)|f_{n}|^{2}.
\end{equation}
From  Theorem I, we know that the Kernel are orthogonal 
to the unitary subspace, which implies that the second term in above 
$\bar{E}(t)$ is zero. The third term can be treated by RUT
as
\begin{equation}
\bar{E}^{1}(t)=\left[\langle\Psi^{1}(t)|U^{\dagger}\right]\left[UH(t)
U^{\dagger}\right]\left[U|\Psi^{1}(t)\rangle\right].
\end{equation}
A simple calculation gives
\begin{equation}
\bar{E}^{1}(t)=\sum\limits_{k=0}^{\infty}\left\{
|\alpha^{+}_{k}|^{2}\left[h_{k}^{2}\lambda-(1+h_{k}^{2})\omega_{k}^{+}
\right]+
|\alpha^{-}_{k}|^{2}\left[\lambda h_{k}^{-2}-
(1+h_{k}^{-2})\omega_{k}^{-}\right]-\lambda \xi_{k}(t)\right\}.
\end{equation}
The variation of the mean energy $\bar{E}(t)$ with time is still
dominated by $ \xi_{k}(t)$, which measure the energy exchange 
of the radiation-matter system with the external source. 

From the above equations (62) and (71), we abstract  
the following interesting relation between the mean energy 
and the atomic inversion,
\begin{equation}
\frac{d \bar{E}}{dt}=-\frac{\lambda}{2} \frac{d\bar{\sigma}_{3}}{dt}.
\end{equation}
This relation expresses such a distinct property of this model
that by measuring the change of energy with time,
we can obtain the information such as the dependence
of the atomic inversion and the mean photon number on time. 
However, this property does not mean that 
the atomic inversion is caused by the external source. In fact, when
$\lambda =0$, $\bar{E}$ turns out to be independent of time, 
but the atomic inversion shown in Eq. (67) still exists  generally.\\

\begin{center}
\subsection*{ 3.2.~ Three-level atoms interacting with one-mode
cavity }
\end{center}

We now  concentrate on the three-level
atom-radiation systems. The three-level atoms are  classified
by the configurations of their energy levels. Generally,
there are three kinds of configurations
$\Lambda, ~V$ and $\equiv$, as shown in Fig. 2,
where the atoms with $\Lambda$-configuration  energy level
have been widely used in 
the subjects such as the coherent population trapping,
laser cooling  \cite{trap}, lasing without inversion \cite{inver},
and electromagnetically induced transparency (EIT) \cite{EIT}.

We know that the exact solvability of various two-level JC models
is attributed to a unified formulation of JC models
by a deformed oscillator algebra \cite{bona}, or by a 
$su(2)$ structure \cite{yu}. However, there  is not a unified 
solvable formulation to these three-level systems. 
Start from this subsection, we will show that by
the RUT method, we can solve various 
three-level models mentioned above.
To avoid the paper being too long, almost all the physical properties
associated with the solutions are not discussed in this paper.

In this subsection, we consider a  system of three-level 
atoms with energies
$\varepsilon_{1}$, $\varepsilon_{2}$ and 
$\varepsilon_{3}$ in the $\Lambda$-configuration,
which interact with a one-mode cavity field as shown in Fig. 2.a.
The interaction is considered to be 
density [$\rho_{1}(N),~ \rho_{2}(N)$]-dependent, and
in the multi-photon ($l_{1},~l_{2}$) form. 
Under the rotating-wave approximation, the Hamiltonian of the 
system is 
written as
\begin{equation}
H=\sum\limits_{i=1}^{3} \varepsilon_{i} S_{ii}+\omega a^{\dagger} a+
[a^{\dagger l_{1}}\rho_{1}(N)S_{12}+\rho_{2}(N)a^{l_{2}}S_{23}+h.c.],
\end{equation}
where $\omega$ is the frequency of the radiation field,
$S_{ij}$  are the  atomic operators given by $S_{ij}=
|i\rangle\langle j|$, $i,j=1,2,3$. We notice that various three-level
Hamiltonians with one-mode cavity in the 
literature are covered by this general one.
To solve this Hamiltonian, we now introduce the 
following  operator $U$:
\begin{equation}
U=\left (
\begin{array}{ccc}
F_{l_{1}} & 0 & 0\\
0 & 1 & 0\\
0 & 0 & F_{l_{2}}
\end{array}
\right ),
\end{equation}
where $ F_{l_{1}}$ and $ F_{l_{2}}$ are two phase operators. 
From  Theorem VI, we know that the operator $U$ belongs to RUT. 
The Kernel of $U$ is
\begin{equation} 
K=\left\{|\psi^{0}_{1}(k_{1})\rangle=
\left(
\begin{array}{c}
|k_{1}\rangle\\
0\\
0
\end{array}
\right), ~~|\psi^{0}_{2}(k_{2})\rangle=
\left(
\begin{array}{c}
0\\
0\\
|k_{2}\rangle
\end{array}
\right),~~~k_{1}<l_{1},~ k_{2}< l_{2} \right\}.
\end{equation}
 Theorem IV
tells us that if this Kernel is isomorphic with to a subset of the
eigenkets of $H$, then the operator $U$ can be applied to $H$. 
It is easily checked that  above $|\psi^{0}_{1}(k_{1})\rangle$
and $|\psi^{0}_{2}(k_{2})\rangle$ happen to be
the eigenkets of $H$.
The eigenvalues corresponding to these eigenkets are
\begin{equation}
\left\{
\begin{array}{l}
E_{1}^{0}(k_{1})=\varepsilon_{1}+\omega k_{1},\\[.15in]
E_{2}^{0}(k_{2})=\varepsilon_{3}+\omega k_{2},
\end{array}
\right.
\end{equation}
where we have assumed that the energy spectrum is  
nondegenerate between two
sets $\{|\psi^{0}_{1}(k_{1})\rangle\}$ and 
$\{|\psi^{0}_{2}(k_{2})\rangle\}$. 
These results indicate that the operator $U$ 
is covered by  Theorem IV, 
and can be applied to the remaining eigenket set 
of $H$ besides those in
Eq. (75). 

Let $H'=UHU^{\dagger}$,  a direct calculation gives
\begin{equation}
H'=\frac{1}{3}(\varepsilon_{1}+\varepsilon_{2}+\varepsilon_{3})
+\omega \left(N+\frac{l_{1}+l_{1}}{3}\right)+H_{1},
\end{equation}
where $H_{1}$ is\\
\begin{equation}
H_{1}=\left (
\begin{array}{ccc}
h_{1} & g_{1}(N) & 0\\
g_{1}(N) & -(h_{1}+h_{2}) & g_{2}(N)\\
0 & g_{2}(N) & h_{2}
\end{array}
\right ),
\end{equation}\\
where  $h_{1}$ and  $h_{2}$ are two constants:
$h_{1}=(2\varepsilon_{1}-\varepsilon_{2}-
\varepsilon_{3})/3+(2l_{1}-l_{2})\omega/3$,~ 
$h_{2}=(2\varepsilon_{3}-\varepsilon_{1}-
\varepsilon_{2})/3+(2l_{2}-l_{1})\omega/3$; and
$g_{1}(N )= \rho_{1}(N) [( N+l_{1})!/N!]^{\frac{1}{2}}$,
$g_{2}(N )= \rho_{2}(N) 
[( N+l_{2})!/N!]^{\frac{1}{2}}$. Notice 
that $H_{1}$ is constructed to be a traceless operator matrix,
this form is easier to compute its eigenvalues. 

The above results show that 
in the transformed frame, the Hamiltonian of the three-level 
matter-radiation interaction system
becomes a $3\times 3$ matrix which depends on
the photon number $N$ only. For the ket vector with  a single 
photon number, this Hamiltonian simply describes the usual three-
level
stationary system, which can be solved exactly. 
We note that in the transformed 
frame, the transitions induced by dipole occur only between the 
states
containing the same photon number. This picture is similar that
in the  two-level system. 

Assuming the eigenket of $H_{1}$ as
\begin{equation}
|\Psi(n)\rangle^{\prime}=
\left(
\begin{array}{c}
\beta_{1}\\
\beta_{2}\\
\beta_{3}
\end{array}
\right)\bigotimes |n\rangle.
\end{equation} 
where $\beta_{i}\equiv \beta_{i}(n)$, $i=1,2,3$.
The eigenvalue $\lambda$ of $H_{1}$ is determined by the 
equation
~$\det|H_{1}-\lambda |=0$, that is
\begin{equation}
\lambda^{3}+\lambda p+q=0,
\end{equation}
where
\begin{equation}\left\{
\begin{array}{l}
p=-\left[g_{1}^{2}(n )+g_{2}^{2}(n )+h_{1}^{2}+h_{2}^{2}+h_{1}h_{2}
\right],\\[.15in]
q=h_{1}g_{1}^{2}(n )+h_{2}g_{1}^{2}(n )+(h_{1}+h_{2})h_{1}h_{2}.
\end{array}
\right.
\end{equation}
The solutions of Eq. (80) are given by \cite{bulitin},
\begin{equation}\left\{
\begin{array}{l}
\lambda_{1}(n)=A_{+}+A_{-}, \\[.15in]
\lambda_{2,3}(n)=
\frac{1}{2}(A_{+}+A_{-})\pm  \frac{\sqrt{3}~i}{2}(A_{+}-A_{-}),
\end{array}
\right.
\end{equation}
where $A_{\pm} \equiv A_{\pm}(n)$,
\begin{equation}
A_{\pm}=\sqrt[3]{\frac{1}{2}q \pm \left(\frac{1}{27}p^{3}+
\frac{1}{4}q^{2}\right)^{1/2}}.
\end{equation}
Combining these $\lambda_{i}$ with
the diagonal term in
$H'$, we obtain the eigenvalues of the Hamiltonian as 
\begin{equation}\left\{
\begin{array}{l}
E_{1}(n)=\frac{1}{3}(\varepsilon_{1}+\varepsilon_{2}+\varepsilon_{3}
)
+\omega \left(n+\frac{l_{1}+l_{1}}{3}\right)
+\lambda_{1}, \\[.15in]
E_{2,3}(n)=\frac{1}{3}(\varepsilon_{1}+\varepsilon_{2}+\varepsilon_{
3})
+\omega \left(n+\frac{l_{1}+l_{1}}{3}\right)+
\lambda_{2,3}.
\end{array}
\right.
\end{equation}
We assume that the energy spectrum is nondegenerate,
then, the eigenkets of $H'$ corresponding to the above eigenvalues 
are
orthogonal mutually, which are obtained to be \\
\begin{equation}
|\Psi_{i}(n)\rangle^{\prime}=
\frac{1}{\zeta_{i}(n)}\left(
\begin{array}{c}
\frac{g_{1}(n)}{\lambda_{i}-h_{1}}\\
1\\
\frac{g_{2}(n)}{\lambda_{i}-h_{2}}
\end{array}
\right)\bigotimes |n\rangle,~~~~~i=1,2,3,
\end{equation} \\
where $\zeta_{i}(n)$ are the normalization factors:
$\zeta_{i}(n)=\left[1+\left(\frac{g_{1}(n)}{\lambda_{i}-
h_{1}}\right)^{2}
+\left(\frac{g_{2}(n)}{\lambda_{i}-h_{2}}\right)^{2}\right]^{1/2}$.
The above solutions of the three-level system
with $\Lambda$-configuration is in the transformed frame. From  
Theorem
IV, we can directly obtain the eigenkets in the original frame  $H$
by the equation: $|\Psi_{i}(n)\rangle=
U^{\dagger}|\Psi_{i}(n)\rangle^{\prime}$, that is\\
\begin{equation}
|\Psi_{i}(n)\rangle=
\frac{1}{\zeta_{i}(n)}\left(
\begin{array}{c}
\left[\frac{g_{1}(n)}{\lambda_{i}-h_{1}}\right]|n+l_{1}\rangle\\
|n\rangle\\
\left[\frac{g_{2}(n)}{\lambda_{i}-h_{1}}\right]|n+l_{2}\rangle
\end{array}
\right),~~~~~i=1,2,3.
\end{equation}\\ 
For an arbitrary initial state $|\varphi(0)\rangle$,
the evolution of the state with time becomes
\begin{equation}
|\varphi(t)\rangle=\sum\limits_{k=0}^{l_{1}-1}\sum\limits_{i=1}^{2}
C_{i}(k)e^{-iE_{i}^{0}(k)t}|\psi^{0}_{i}(k)\rangle+
\sum\limits_{n=0}^{\infty}\sum\limits_{j=1}^{3}
D_{j}(n)e^{-iE_{j}(n)t}|\Psi_{j}(n)\rangle,
\end{equation}
where $C_{i}(k)=\langle\psi_{i}^{0}(k)|\varphi(0)\rangle $, 
and $D_{j}(n)=\langle\Psi_{j}(n)|\varphi(0)\rangle $.

Now we conclude that the total eigenstate set of 
the three-level atom-radiation system is constituted by two
subsets as in Eqs. (75) and (86), 
where the subset in Eq. (75) is the kernel of
$U$, but for the other subset Eq. (86), $U$ acts as a unitary operator.

We recall that for the two-level JC model, there generally 
exists a relation
$H'\sim\sigma\cdot B$.
However, for the above three-level system, from the expression of 
$H'$, we know that the system can
not be taken as a spin-$1$ particle interacting with an external 
magnetic 
field, that is, the relation such as $H'\sim S\cdot B$
does not exist here. Therefore, there is no concept
of spin in the $\Lambda$-configuration system,
and many methods developed for
the two-level JC model do not fit for the three-level case.
However, for the RUT method, one may notice that the procedure in 
solving
the three-level model is exactly the same as the procedure
presented in Ref. [9] in solving the two-level JC model.

To investigate the dynamics of the above system, 
we usually use the method of density matrix.  Supposing at $t=0$
the density operator as
\begin{equation}
\rho(0)=\sum\limits_{k_{1},k_{2}=0}^{\infty}|\varphi(k_{1})\rangle
\langle \varphi'(k_{2})|.
\end{equation}
Using Eq. (87), one can directly 
obtain the evolution of the density operator with time
by the equation $\rho(t)=e^{-iHt}\rho(0)e^{iHt}$. With 
$\rho(t)$, one can calculate some quantities
such as the atomic inversion, population trapping in this model.
We now choose such a special initial state that
the atoms are in the level $i=1$, and the photon state is 
represented by the coherent state. Then  
 $\rho(0)$ is written into
\begin{equation}
\rho(0)=\sum\limits_{k_{1},k_{2}=0}^{\infty}\frac{e^{-|z|^{2}}z^{k_{1}} 
z^{* k_{2}}}{\sqrt{(k_{1}! k_{2}!)}}|k_{1}\rangle
\langle k_{2}|S_{11}.
\end{equation}
 Make use of Eq. (87), we obtain
\begin{equation}
\rho(t)=|Z(t)\rangle\langle Z(t)|,
\end{equation}
where
\begin{equation}
|Z(t)\rangle=\sum\limits_{k=0}^{l_{1}-1}
\frac{e^{-\frac{|z|^{2}}{2}}z^{k}} 
{\sqrt{k!}}
e^{-iE_{1}^{0}(k)t}|\psi^{0}_{1}(k)\rangle+
\sum\limits_{n=0}^{\infty}\sum\limits_{j=1}^{3}
\frac{e^{-\frac{|z|^{2}}{2}}z^{n+l_{1}}g_{1}(n)e^{-iE_{j}(n)t}}
{\sqrt{(n+l_{1})!}\zeta_{j}(n)[\lambda_{j}(n)-h_{1}]}
|\Psi_{j}(n)\rangle.
\end{equation}\\
Using $\rho(t)$, we can obtain the probability of finding the atom
in the state $i=2$ which is initially in the state $i=1$: 
$ P_{1\rightarrow 2}(t)=\sum\limits_{k=0}^{\infty}\langle i=2, k|
\rho(t)|k, i=2\rangle$,
\begin{equation}
P_{1\rightarrow 2}(t)=\sum_{k=0}^{\infty}
\frac{e^{-x}x^{k+l_{1}}}{ (k+l_{1})!}\left|\sum\limits_{i=1}^{3}
\frac{g_{1}(k)~e^{-i\lambda_{i}(k)t}}
{\zeta_{i}^{2}(k)[\lambda_{i}(k)-h_{1}]}~
\right|^{2},
\end{equation}
where $x=|z|^{2}$. 
We know that the above transition $(1\leftrightarrow 2)$ is induced
by dipole. A more interesting quantity is $P_{1\rightarrow 3}(t)$,
which is induced by two dipoles: $(1\leftrightarrow 2)$
and $(2\leftrightarrow 3)$,
\begin{equation}
P_{1\rightarrow 3}(t)=\sum_{k}
\frac{e^{-x}x^{k+l_{1}}}{ (k+l_{1})!}\left|\sum\limits_{i=1}^{3}
\frac{g_{1}(k)g_{2}(k)~e^{-i\lambda_{i}(k)t}}
{\zeta_{i}^{2}(k)[\lambda_{i}(k)-h_{1}]
[\lambda_{i}(k)-h_{2}]}~
\right|^{2}.
\end{equation}
The statistics of the photon number at the state $i=3$ is 
obtained to be
\begin{equation}
\bar{n}_{3}(t)=\sum_{k}
\frac{e^{-x}x^{k+l_{1}}}{ (k+l_{1})!}\left|\sum\limits_{i=1}^{3}
\frac{g_{1}(k)g_{2}(k)~(k+l_{2})~e^{-i\lambda_{i}(k)t}}
{\zeta_{i}^{2}(k)[\lambda_{i}(k)-h_{1}]
[\lambda_{i}(k)-h_{2}]}~
\right|^{2}.
\end{equation}
The atomic inversion of the system under the special initial 
state is $\bar{S}_{3}(t)\equiv \langle S_{3} \rangle$, which
is explicitly given by
\begin{equation}
\bar{S}_{3}(t)=\sum_{k}
\frac{e^{-x}x^{k+l_{1}}}{ (k+l_{1})!}
\left|\sum\limits_{i=1}^{3}
\frac{g_{1}^{2}(k)~e^{-i\lambda_{i}(k)t}}
{\zeta_{i}^{2}(k)[\lambda_{i}(k)-h_{1}]^{2}}~
\right|^{2}-P_{1\rightarrow 3}(t).
\end{equation}\\

\begin{center}
\subsection*{ 3.3. ~ Three-level atoms interacting with a two-mode
cavity field }
\end{center}

In this subsection, we  consider a system of
three-level atoms with energies $\varepsilon_{1}, ~
\varepsilon_{2}$ and $\varepsilon_{3}$, which 
 interacts with a two-mode cavity field:
a pump mode ($a_{1}, ~a_{1}^{\dagger}$) 
of frequency $\omega_{1}$
and a Stokes mode ($a_{2}, ~a_{2}^{\dagger}$) of frequency 
$\omega_{2}$  as shown in Fig. 2.a. The interactions
are generalized to be multiphoton ($l_{1},~l_{2}$) forms, and 
dependent 
on densities $\rho_{1}(N_{1}),~ \rho_{2}(N_{2})$, respectively.
Therefore, the Hamiltonian is 
\begin{equation}
H=H_{0}+H_{1},
\end{equation}
where
\begin{equation}
\begin{array}{l}
H_{0}=\sum\limits_{i=1}^{3} \varepsilon_{i} S_{ii}+
\omega_{1} a_{1}^{\dagger} a_{1}+\omega_{2} a_{2}^{\dagger} a_{2}, 
\\[.15in]
H_{1}=a_{1}^{\dagger l_{1}}\rho_{1}(N_{1})S_{12}+
\rho_{2}(N_{2})a_{2}^{l_{2}}S_{23}+h.c..
\end{array}
\end{equation}
This Hamiltonian covers various special cases in the literature.
In the above Subsection {\bf 3.2} we have shown that the solution of 
three-level case is much more complicated than that of the 
two-level case. To avoid the complication, 
one usually treats $H$ by the perturbation theory 
in the interaction picture. 
Especially in the case  that the  detunings of 
the two modes are very large, the upper level
can be eliminated adiabatically from the three-level system,
then the system is reduced into a simple  two-level case with two
quantized modes \cite{eberly}, where the Stark shift terms appear 
and 
give arise to some interesting physical effects \cite{scully}.  

We divide this subsection into two parts.
In the first part {\bf  3.3.1},  we 
consider the case that the detunings of the system 
are enough large, then by constructing a unitary transformation,
we separate the upper level 
from the system to the first order approximation, the remaining 
two low levels are still treated by the RUT method. We further
discuss the relation between the so-called ``dressed" states
and ``bare" states from the viewpoint of transformation.
In the second part {\bf  3.3.2}, we use the same RUT method to 
solve the system exactly.\\

\begin{center}
\subsection*{ 3.3.1. ~Approximate treatment in the 
case of large detunings }
\end{center}

In the case that the detunings of 
two modes are very large, we accordingly 
introduce the following unitary transformation $Q$,
\begin{equation}
Q=\exp(X),
\end{equation}
where
\begin{equation}
X=\left[-\delta_{1}a_{1}^{\dagger l_{1}}\rho_{1}(N_{1})S_{12}+
\delta_{2}\rho_{2}(N_{2})a_{2}^{l_{2}}S_{23}\right]-h.c.,
\end{equation}
where $\delta_{1}$ and $\delta_{2}$ 
are two parameters to be determined. 
Under  $Q$, the Hamiltonian Eq. (96)
is transformed into 
$H'=QHQ^{\dagger}$. Using the Baker-Hausdorff formula, $H'$ 
becomes
\begin{equation}
H'=H_{0}+ H_{1}+[X, H_{0}]+[X, H_{1}]+\frac{1}{2}[X,[X, H_{0}]]
+\frac{1}{2}[X,[X, H_{1}]]+\cdots.
\end{equation}
If we choose $\delta_{1}=\Delta_{1}^{-1}
\equiv(\varepsilon_{2}-\varepsilon_{1}-\omega_{1}l_{1})^{-1},~
\delta_{2}=\Delta_{2}^{-1}\equiv(\varepsilon_{2}-\varepsilon_{3}-
\omega_{2}l_{2})^{-1}$, 
where $\Delta_{1}$
and $\Delta_{2}$ are two detunings, then
\begin{equation}
[X, H_{0}]=-H_{1}.
\end{equation}
For the large $\Delta_{i},~i=1, 2$, we expand $H'$ 
up to the first order of $1/\Delta_{i}$ as
\begin{equation}
H'\approx H_{0}+[X, H_{1}]+\frac{1}{2}[X,[X, H_{0}]]=
H_{0}+\frac{1}{2}[X, H_{1}],
\end{equation}
that is\\
\begin{eqnarray}
H' & = & 
\sum_{i=1}^{2}\omega_{i}a^{\dagger}_{i}a_{i}+\left[\varepsilon_{1}-
\frac{1}{\Delta_{1}}\rho^{2}_{1}(N_{1}-l_{1})a_{1}^{\dagger l_{1}}
a_{1}^{l_{1}}\right]S_{11}+\left[\varepsilon_{3}-
\frac{1}{\Delta_{2}}\rho^{2}_{2}(N_{2}-l_{2})a_{2}^{\dagger l_{2}}
a_{2}^{l_{2}}\right]S_{33} \nonumber\\[.15in]
& & 
-\left\{\left[\frac{1}{2}(\frac{1}{\Delta_{1}}+\frac{1}{\Delta_{2}})
\rho_{1}(N_{1}-l_{1})\rho_{2}(N_{2})a_{1}^{\dagger l_{1}}
a_{2}^{l_{2}}\right]S_{13}+h.c.\right\}\nonumber\\[.15in]
& & +\left[
\varepsilon_{2}+\frac{1}{\Delta_{1}}\rho^{2}_{1}(N_{1})a_{1}^{l_{1}}
a_{1}^{\dagger l_{1}}+
\frac{1}{\Delta_{2}}\rho^{2}_{2}(N_{2})a_{2}^{l_{2}}
a_{2}^{\dagger l_{2}}\right]S_{22}.\label{eq3}
\end{eqnarray}\\

We have two comments on the above results:

i. The Hamiltonian $H'$ indicates that if the atomic states are in the 
form
\begin{equation}
|\varphi\rangle'=
\left(
\begin{array}{c}
|\varphi_{1}\rangle\\
0\\
|\varphi_{3}\rangle
\end{array}
\right),
\end{equation} 
the term about $S_{22}$ in $H'$ does not contribute. 
Then the system is reduced into a two-level atomic system.
The two terms $-\frac{1}{\Delta_{1}}\rho^{2}_{1}(N_{1}-l_{1})
a_{1}^{\dagger l_{1}}
a_{1}^{l_{1}}S_{11}$ and $-
\frac{1}{\Delta_{2}}\rho^{2}_{2}(N_{2}-l_{2})a_{2}^{\dagger l_{2}}
a_{2}^{l_{2}}S_{33}$ appearing in the Hamiltonian are 
called Stark effect terms, which are in terms of the photon numbers.

ii. We regard the operators in the original frame $H$
as ``bare" operators, and the atomic states as ``bare" states,
then the operators in the transformed frame $H'$ are ``dressed"
operators, and the atomic states are ``dressed" states. The state
Eq. (104) is in fact a ``dressed" state.
We know that all the states  prepared in experiment
are ``bare" states. Therefore, turned back to 
experiment, the ``bare" state corresponding to Eq. (104) is
\begin{equation}
|\varphi\rangle=e^{-X}|\varphi\rangle'\approx
\left(
\begin{array}{c}
|\varphi_{1}\rangle\\
\frac{\rho_{1}(N_{1})}{\Delta_{1}}a_{1}^{l_{1}}|\varphi_{1}\rangle+
\frac{\rho_{2}(N_{2})}{\Delta_{2}}a_{2}^{l_{2}}|\varphi_{3}\rangle\\
|\psi_{3}\rangle
\end{array}
\right).
\end{equation} 
This result shows that the atomic state $i=2$ still contributes
in the original system. 

We now solve the Hamiltonian $H'$ by the RUT method developed 
above.
Consider an operator $V$  
\begin{equation}
V=\left (
\begin{array}{ccc}
F_{l_{1}}(1)  & 0 & 0\\
0 & 1 & 0\\
0 & 0 & F_{l_{2}}(2)
\end{array}
\right),
\end{equation}
where $F_{l_{1}}(1)=\left[a_{1}^{l_{1}}
a_{1}^{\dagger l_{1} } \right]^{-1/2}a_{1}^{l_{1}}$, 
$F_{l_{2}}(2)=\left[a_{2}^{l_{2}}
a_{2}^{\dagger l_{2} }\right]^{-1/2}a_{2}^{l_{2}}$ are two
phase operators. Thus $V$ belongs to RUT, 
and its Kernel is
\begin{equation} 
K=\left\{|\psi^{0}_{1}(k_{1}, k_{2})\rangle=
\left(
\begin{array}{c}
|k_{1}, k_{2}\rangle\\
0\\
0
\end{array}
\right),~~
|\psi^{0}_{2}( k'_{1}, k'_{2})\rangle=
\left(
\begin{array}{c}
0\\
0\\
|k'_{1}, k'_{2}\rangle
\end{array}
\right),~~~k_{1}<l_{1},~ k'_{2}< l_{2} \right\}.
\end{equation}\\
We obtain that $|\psi^{0}_{1}(k_{1}, k_{2})\rangle$ 
are the eigenkets of $H'$,  the eigenvalues are
\begin{equation}
E_{1}^{0}(k_{1}, 
k_{2})=\varepsilon_{1}+\omega_{1}k_{1}+\omega_{2}k_{2};
\end{equation}
$|\psi_{2}( k'_{1}, k'_{2})\rangle$ are the eigenkets of $H'$
with the eigenvalues 
\begin{equation}
E_{2}^{0}( k'_{1}, 
k'_{2})=\varepsilon_{3}+\omega_{1}k'_{1}+\omega_{2}k'_{2}.
\end{equation}
Where we have assumed that there is no degeneracy between
 $E_{1}^{0}(k_{1}, k_{2})$ and $E_{2}^{0}( k'_{1}, k'_{2})$. 
These results indicate that  $V$ is covered by  Theorem IV, 
and can be applied to $H'$ to obtain the remaining subspace besides
Eq. (107). We make  the following transformation,
\begin{equation}
H''=VH'V^{\dagger}, 
\end{equation}
that is
\begin{eqnarray}
H''& =&
B_{0}(N_{1}, N_{2})(S_{11}+S_{33})+B_{3}
(N_{1}, N_{2})(S_{11}-S_{33})\nonumber\\
& &+B_{1}(N_{1}, N_{2})
(S_{13}+S_{31})+B_{4}(N_{1}, N_{2})S_{22},
\end{eqnarray}
where $B_{i}(N_{1}, N_{2}),~i=0,1,3,4,$ are given by\\
\begin{equation}
\left\{
\begin{array}{l}
B_{0}(N_{1}, N_{2})=\frac{1}{2}(\varepsilon_{1}+\varepsilon_{3})+
\sum\limits_{i=1}^{2}\left[\omega_{i}(N_{i}+\frac{l_{i}}{2})-
\frac{1}{2\Delta_{i}}g_{i}^{2}(N_{i})\right],\\[.15in]
B_{1}(N_{1}, N_{2})=-\frac{1}{2}
(\frac{1}{\Delta_{1}}+\frac{1}{\Delta_{2}})
g_{1}(N_{1})g_{2}(N_{2}),\\[.15in]
B_{3}(N_{1}, N_{2})=\frac{1}{2}(\varepsilon_{1}-
\varepsilon_{3})+\frac{1}{2}
\left[\omega_{1}l_{1}-\omega_{2}l_{2}+
\frac{g_{1}^{2}(N_{1})}{\Delta_{1}}-
\frac{g_{2}^{2}(N_{2})}{\Delta_{2}}\right],\\[.15in]
B_{4}(N_{1}, N_{2})=\varepsilon_{2}+\sum\limits_{i=1}^{2}
\left[\omega_{i}N_{i}
+\frac{1}{\Delta_{i}}g_{i}^{2}(N_{i})\right],
\end{array}
\right.
\end{equation}\\
where $g_{i}(N_{i} )= \rho_{i}(N_{i}) 
[( N_{i}+l_{i})!/N_{i}!]^{\frac{1}{2}},~ i=1,2$.
The Hamiltonian $H''$ is a function of $N_{1}$ and $N_{2}$, and
composed of two independent parts: 
$H''_{1}\equiv B_{0}(N_{1}, N_{2})(S_{11}+S_{33}) +
B_{3}(N_{1}, N_{2})(S_{11}-S_{33})+B_{1}(N_{1}, N_{2})
(S_{13}+S_{31})$ 
is a two-level system containing
a two-mode cavity; the other part is  
$H''_{2}\equiv B_{4}(N_{1}, N_{2})S_{22}$. The eigenkets 
of $H''_{1}$ are directly written as
\begin{equation}
|\Psi^{+}(k_{1}, k_{2})\rangle''=
\left(
\begin{array}{c}
\cos\frac{\theta}{2}|k_{1}, k_{2}\rangle\\
0\\
\sin\frac{\theta}{2}|k_{1}, k_{2}\rangle
\end{array}
\right),~~~
|\Psi^{-}( k_{1}, k_{2})\rangle''=
\left(
\begin{array}{c}
-\sin\frac{\theta}{2}|k_{1}, k_{2}\rangle\\
0\\
\cos\frac{\theta}{2}|k_{1}, k_{2}\rangle
\end{array}
\right),
\end{equation} 
where $\theta\equiv \theta (k_{1}, k_{2})=
\cos^{-1}\left\{ B_{3}(k_{1}, k_{2})/\left[B_{1}^{2}(k_{1}, k_{2})+
B_{3}^{2}(k_{1}, k_{2})\right]^{1/2}\right\}$.
The eigenvalues corresponding to these two kets are respectively
\begin{equation}
E^{\pm}(k_{1}, k_{2})=B_{0}(k_{1}, k_{2})
\pm \left[B_{1}^{2}(k_{1}, k_{2})+
B_{3}^{2}(k_{1}, k_{2})\right]^{1/2}.
\end{equation}
The third eigenket set of $H''$ is determined by $H_{2}''$,
\begin{equation}
|\Psi_{3}(k_{1}, k_{2})\rangle''=
\left(
\begin{array}{c}
0\\
|k_{1}, k_{2}\rangle\\
0
\end{array}
\right),
\end{equation}  
the eigenvalues are simply
\begin{equation}
E_{3}(k_{1}, k_{2})=B_{4}(k_{1}, k_{2}).
\end{equation}
Using $V^{\dagger}$, we can turn these ket vectors to  
the original frame $H'$. $|\Psi_{3}(k_{1}, k_{2})\rangle''$ 
are invariant under $V^{\dagger}$. But
$|\Psi^{\pm}(k_{1}, k_{2})\rangle''$ turn out to be
\begin{equation} 
|\Psi^{+}(k_{1}, k_{2})\rangle'=
\left(
\begin{array}{c}
\cos\frac{\theta}{2}|k_{1}+l_{1}, k_{2}\rangle\\
0\\
\sin\frac{\theta}{2}|k_{1}, k_{2}+l_{2}\rangle
\end{array}
\right),~~~
|\Psi^{-}( k_{1}, k_{2})\rangle'=
\left(
\begin{array}{c}
-\sin\frac{\theta}{2}|k_{1}+l_{1}, k_{2}\rangle\\
0\\
\cos\frac{\theta}{2}|k_{1}, k_{2}+l_{2}\rangle
\end{array}
\right).
\end{equation}\\
The results (114) and (117) indicate the existence of Rabi frequency
\begin{equation} 
\omega(k_{1}, k_{2})=\left[B_{1}^{2}(k_{1}, k_{2})+
B_{3}^{2}(k_{1}, k_{2})\right]^{1/2}
\end{equation}
in the system $H'$. Therefore, some interesting effects 
appearing in the usual two-level system can also appear here.
However, the above comment (ii) has indicated that these ket 
vectors (117) are in fact the ``dressed"  ket vectors, 
induced by the unitary transformation $Q$ as Eq. (98). 
For the viewpoint of experiment, it is meaningful to 
find these ket vectors in the ``bare" form. Up to the first 
order approximation, we can use Eq. (105),  and find that 
$|\Psi_{3}(k_{1}, k_{2})\rangle''$ maintain the same form, but
the vectors $|\Psi^{\pm}(k_{1}, k_{2})\rangle'$ turn out to be 
\begin{equation} 
|\Psi^{+}(k_{1}, k_{2})\rangle=
\left(
\begin{array}{c}
\cos\frac{\theta}{2}|k_{1}+l_{1}, k_{2}\rangle\\
\chi_{1}|k_{1}, k_{2}\rangle\\
\sin\frac{\theta}{2}|k_{1}, k_{2}+l_{2}\rangle
\end{array}
\right),~~~
|\Psi^{-}( k_{1}, k_{2})\rangle=
\left(
\begin{array}{c}
-\sin\frac{\theta}{2}|k_{1}+l_{1}, k_{2}\rangle\\
\chi_{2}|k_{1}, k_{2}\rangle\\
\cos\frac{\theta}{2}|k_{1}, k_{2}+l_{2}\rangle
\end{array}
\right),
\end{equation}
where 
\begin{equation}
\left\{
\begin{array}{l}
\chi_{1}=\frac{\rho_{1}(k_{1})}{
\Delta_{1}}\sqrt{\frac{(k_{1}+l_{1})!}{k_{1}}}
\cos\frac{\theta}{2}+\frac{\rho_{1}(k_{2})}{
\Delta_{2}}\sqrt{\frac{(k_{2}+l_{2})!}{k_{2}}}
\sin\frac{\theta}{2}, \\[.15in]
\chi_{2}=-\frac{\rho_{1}(k_{1})}{
\Delta_{1}}\sqrt{\frac{(k_{1}+l_{1})!}{k_{1}}}
\sin\frac{\theta}{2}+\frac{\rho_{1}(k_{2})}{
\Delta_{2}}\sqrt{\frac{(k_{2}+l_{2})!}{k_{2}}}
\cos\frac{\theta}{2}.
\end{array}
\right.
\end{equation}
The normalization factors are not included in the above 
ket vectors.
Now we can conclude that only for the ket vectors as
Eq. (119), the Rabi oscillation can appear in experiment.

Recall that $H'$ results from the first order approximation 
of $H$ under the condition that $\langle\frac{g_{1}^{2}(N_{1})}
{\Delta_{1}}\rangle,
\langle\frac{g_{2}^{2}(N_{2})}{\Delta_{2}}\rangle 
\ll~\Delta_{1},~\Delta_{2}$.
We now consider the following two special cases
to reduce the above eigenvalues Eq. (114):

First, for the case that is often used in the literature:
\begin{equation}
\Delta_{1}=\Delta_{2}=\Delta,
\end{equation}
these $E^{\pm}(k_{1}, k_{2})$  are reduced into
\begin{equation}
E^{\pm}(k_{1}, k_{2})=\frac{1}{2}(\varepsilon_{1}+\varepsilon_{3})+
\sum\limits_{i=1}^{2}\left[\omega_{i}(k_{i}+\frac{l_{i}}{2})-
\frac{1}{2\Delta}g_{i}^{2}(k_{i})\right]\pm
\left|\frac{g_{1}^{2}(k_{1})-g_{2}^{2}(k_{2})}{2\Delta}\right|.
\end{equation}

The second case is contrary to Eq. (121) that
\begin{equation}
\langle\frac{g_{1}^{2}(N_{1})}{\Delta_{1}}\rangle,
\langle\frac{g_{2}^{2}(N_{2})}{\Delta_{2}}\rangle
\ll|\Delta_{1}-\Delta_{2}|,
\end{equation}
we may accordingly reduce  $E^{\pm}(k_{1}, k_{2})$ 
into the following form
\begin{eqnarray}
E^{\pm}(k_{1}, 
k_{2})&=&\frac{1}{2}(\varepsilon_{1}+\varepsilon_{3})+
\sum\limits_{i=1}^{2}\left[\omega_{i}(k_{i}+\frac{l_{i}}{2})-
\frac{1}{2\Delta_{i}}g_{i}^{2}(k_{i})\right]\nonumber\\[.15in]
& & 
\pm \frac{1}{2}\left[\Delta_{2}-\Delta_{1}+
\frac{g_{1}^{2}(k_{1})}{\Delta_{1}}-
\frac{g_{2}^{2}(k_{2})}{\Delta_{2}}+
(\frac{1}{\Delta_{1}}+\frac{1}{\Delta_{2}})
\frac{g_{1}(k_{1})g_{2}(k_{2})}{2(\Delta_{2}-\Delta_{1})}\right].
\label{eq2}
\end{eqnarray}\\

\begin{center}
\subsection*{3.3.2.~ Exact solution}
\end{center}

We know that the approximate solution to $H$ in the above 
subsection
is based on the condition 
\begin{equation}
\langle\frac{g_{1}^{2}(N_{1})}{\Delta_{1}}\rangle,~
\langle\frac{g_{2}^{2}(N_{2})}{\Delta_{2}}\rangle 
\ll \Delta_{1},  ~ \Delta_{2}.
\end{equation}
However, the quantities $g_{i}(N_{i})\equiv \rho_{i}(N_{i}) 
[( N_{i}+l_{i})!/N_{i}!]^{\frac{1}{2}}, ~i=1,~2,$  increase
with the photon numbers $N_{i}$ for certain densities, which means 
that
the condition (125) does not hold for the states of large photon 
numbers .
In this subsection, we will solve
the Hamiltonian (96) exactly.

We employ again the right-unitary operator $V$ in 
Eq. (106) as a transformation.
It is easily checked that the Kernel of $V$ given by Eq. (107) 
is still a set of the eigenkets of $H$. Since the Kernel is 
invariant under the unitary transformation $Q$ [Eq. (98)],  the 
eigenvalues of $H$ corresponding to these eigenkets are still 
the same as those of $H'$  [Eq. (100)]. Based on these results,
we can use $V$ as a unitary transformation to $H$ to
obtain its remaining subspace.  Let 
\begin{equation}
\bar{H}=VHV^{\dagger},
\end{equation}
a direct calculation gives
\begin{equation}
 \bar{H}=\bar{H}_{0}+\bar{H}_{1},
\end{equation}
where
\begin{equation}
\bar{H}_{0}=\sum\limits_{i=1}^{3}\varepsilon_{i}/3+
\sum\limits_{j=1}^{2}\omega_{j}(N_{j}+l_{j}/3),
\end{equation}
and $\bar{H}_{1}$ is a traceless matrix,
\begin{equation}
\bar{H}_{1}=\left (
\begin{array}{ccc}
f_{1} & g_{1}(N_{1}) & 0\\
g_{1}(N_{1}) & -(f_{1}+f_{2}) & g_{2}(N_{2})\\
0 & g_{2}(N_{2}) & f_{2}
\end{array}
\right ),
\end{equation}
where $g_{1}(N_{1})$ and $g_{2}(N_{2})$ are given in Eq. (112),
and 
\begin{equation}
\left\{
\begin{array}{l}
f_{1}=\frac{1}{3}(2\varepsilon_{1}-\varepsilon_{2}-\varepsilon_{3}+
2\omega_{1}l_{1}-\omega_{2}l_{2})=\frac{1}{3}(\Delta_{2}-
2\Delta_{1}),\\[.15in]
f_{2}=\frac{1}{3}(2\varepsilon_{3}-\varepsilon_{1}-\varepsilon_{2}+
2\omega_{2}l_{2}-\omega_{1}l_{1})=\frac{1}{3}(\Delta_{1}-
2\Delta_{2}).
\end{array}
\right.
\end{equation}
The Hamiltonian now becomes a function of photon numbers $N_{1}$
and $N_{2}$, where the creation and annihilation of the photon
in the transitions between the atomic states are erased by the 
transformation $V$. Therefore, in the new frame,
the transitions can only happen
between the atomic states having the same photon number
$N_{1}$ (and  $N_{2}$). For a ket vector with photon numbers 
 $k_{1}$ and $k_{2}$
\begin{equation}
|\Psi(k_{1}, k_{2})\rangle'=\left (
\begin{array}{c}
\beta_{1} \\
\beta_{2}\\
\beta_{3}
\end{array}
\right )\bigotimes |k_{1}, k_{2}\rangle,
\end{equation}
where $\beta_{i}\equiv \beta_{i}(k_{1}, k_{2})$, 
$i=1,2,3$, $\bar{H}$ becomes the usual stationary three-level system,
where the eigenvalues of $\bar{H}_{1}$ are determined by the 
equation
~$\det|\bar{H}_{1}-\lambda |=0$, namely,
\begin{equation}
\lambda^{3}+\lambda p_{1}+q_{1}=0,
\end{equation}
where
\begin{equation}\left\{
\begin{array}{l}
p_{1}=-\left[g_{1}^{2}(k_{1} )+g_{2}^{2}(k_{2})+
f_{1}^{2}+f_{2}^{2}+f_{1}f_{2}
\right],\\[.15in]
q_{1}=f_{1}g_{2}^{2}(k_{2} )+f_{2}g_{1}^{2}(k_{1})+
(f_{1}+f_{2})f_{1}f_{2}.
\end{array}
\right.
\end{equation}
The solutions of this equation are \cite{bulitin},
\begin{equation}
\left\{
\begin{array}{l}
\lambda_{1}\equiv \lambda_{1}(k_{1}, k_{2})=C_{+}+C_{-}, \\[.15in]
\lambda_{2,3}\equiv\lambda_{2,3}(k_{1}, k_{2})=
\frac{1}{2}(C_{+}+C_{-})\pm  \frac{\sqrt{3}~i}{2}(C_{+}-C_{-}),
\end{array}
\right.
\end{equation}
where $C_{\pm} \equiv C_{\pm}(k_{1}, k_{2})$ are given by
\begin{equation}
C_{\pm}=\sqrt[3]{\frac{1}{2}q_{1} \pm \left(\frac{1}{27}p_{1}^{3}+
\frac{1}{4}q_{1}^{2}\right)^{1/2}}.
\end{equation}
Combining the above $\lambda_{i}$ with
the term $\bar{H}_{0}$ in $\bar{H}$, 
we obtain the exact eigenvalues of the three-level
system 
\begin{equation}\left\{
\begin{array}{l}
E_{1}(k_{1}, k_{2})=\sum\limits_{i=1}^{3}\varepsilon_{i}/3+
\sum\limits_{j=1}^{2}\omega_{j}(k_{j}+l_{j}/3)
+\lambda_{1}, \\[.15in]
E_{2,3}(k_{1}, k_{2})=\sum\limits_{i=1}^{3}\varepsilon_{i}/3+
\sum\limits_{j=1}^{2}\omega_{j}(k_{j}+l_{j}/3)+
\lambda_{2,3}.
\end{array}
\right.
\end{equation}

We  assume that there is no degeneracy in the energy spectrum.
Then, the eigenstates of $\bar{H}$ corresponding 
to the above eigenvalues are
orthogonal mutually, which are given by
\begin{equation}
|\Psi_{i}(k_{1}, k_{2})\rangle^{\prime}=
\frac{1}{\xi_{i}}\left(
\begin{array}{c}
\frac{g_{1}}{\lambda_{i}-f_{1}}\\
1\\
\frac{g_{2}}{\lambda_{i}-f_{2}}
\end{array}
\right)\bigotimes |k_{1}, k_{2}\rangle,~~~~~i=1,2,3,
\end{equation} 
where $\xi_{i}$ are the normalization factors:
$\xi_{i}=\left[1+\left(\frac{g_{1}}{\lambda_{i}-f_{1}}\right)^{2}
+\left(\frac{g_{2}}{\lambda_{i}-f_{2}}\right)^{2}\right]^{1/2}$.
Using  Theorem
IV, we obtain the eigenkets in the original frame as:
\begin{equation}
|\Psi_{i}(k_{1}, k_{2})\rangle=
\frac{1}{\xi_{i}}\left(
\begin{array}{c}
\left(\frac{g_{1}}{\lambda_{i}-f_{1}}\right)|k_{1}+l_{1},
k_{2}\rangle\\
|k_{1},
k_{2}\rangle\\
\left(\frac{g_{2}}{\lambda_{i}-f_{2}}\right)|k_{1},
k_{2}+l_{2}\rangle
\end{array}
\right),~~~~~i=1,2,3.
\end{equation}  

Up to now, we have solved exactly the system of 
$\Lambda$-configuration atoms interacting with a two-mode 
cavity field by the RUT method.
One may notice that the above procedure is almost
the same as that in subsection {\bf 3.2},
this means that to the RUT method, that the system 
contains one-mode cavity or two-mode cavity has no difference, 
even though the physical meanings of solutions are quite different.
Using the expression of $f_{1}$ and $f_{2}$, 
one can expand the above solution according to the condition
of large detunings, and compare the result with the approximate
solution obtained in last subsection.
Based on the above solution, one can further study the physical
effects in this model.

In the literature, ones study the  three-level
system where the cavity fields 
change with time, i.e., the nonstationary case. We should
point out that one can follow Theorem V to treat
this nonstationary system, and the procedure is similar to 
that in  above Subsection {\bf 3.1}.\\

\begin{center}
\subsection*{ 3.4. ~Application of RUT to the
 atoms with other configurations }
\end{center}
The above approach concentrates on the system where the atoms 
have
$\Lambda$-configuration energy levels
only. Recently, $V$-configuration atoms were found to
exhibit some new interesting effects \cite{zhu}.
In principle, under the condition that the detunings
are very large, a system of $V$-configuration atoms 
interacting with cavity fields can be treated  by the same way 
as the above approximate treatment to $\Lambda$-configuration 
atoms. 
Unfortunately, for $\equiv$-configuration atomic system, 
this treatment becomes invalid.

However, as we have pointed out that the detunings
are not always very large for various atomic systems. Moreover, 
$g_{i}(N_{i})$ in Eq. (112) always increase with the photon 
number $N_{i}$, but the detunings are invariant with $N_{i}$.
Therefore, the perturbation treatment introduced above is 
valid only to the situation of large detunings 
and low photon number state.  
One may notice that the exact treatment of three-level atomic system
by RUT method is evidently simpler than the perturbation treatment, 
and 
suitable to various situations.
In this subsection, we briefly show how to apply RUT method to the 
systems
of $V$ and $\equiv$-configuration atoms.

We first look at a system of $V$-configuration atoms interacting with
one-mode cavity field shown in Fig. 2.b. 
The generalized Hamiltonian is
\begin{equation}
H=\sum\limits_{i=1}^{3} \varepsilon_{i} S_{ii}+\omega a^{\dagger} a+
[\rho_{1}(N) a^{ l_{1}} S_{12}+a^{\dagger l_{2}} \rho_{2}(N) 
S_{23}+h.c.],
\end{equation}
where we assume $l_{1}\leq l_{2}$.
To solve this Hamiltonian, we here construct an operator $U$\\
\begin{equation}
U=\left (
\begin{array}{ccc}
F_{l_{2}-l_{1}} & 0 & 0\\
0 & F_{l_{2}} & 0\\
0 & 0 & 1
\end{array}
\right ),
\end{equation}\\
where $F_{l_{2}-l_{1}}$ and $ F_{l_{2}}$ are two phaser operators. 
 $U$ belongs to RUT, and the Kernel of $U$ is \\
\begin{equation}
K=\left\{|\psi_{1}^{0}(k_{1})\rangle=
\left(
\begin{array}{c}
|k_{1}\rangle\\
0\\
0
\end{array}
\right), ~~|\psi_{2}^{0}(k_{2})\rangle=
\left(
\begin{array}{c}
0\\
|k_{2}\rangle\\
0
\end{array}
\right),~~~k_{1}<l_{2} -l_{1},~ k_{2}< l_{2} \right\}.
\end{equation}\\
We can take $|\Psi\rangle=\chi_{1}|\psi_{1}^{0}(k_{1})\rangle
+\chi_{2}|\psi_{2}^{0}(k_{2})\rangle$  to check whether
 $|\Psi\rangle$ is the eigenket of $H$. Without difficulty,
we find the following sets of vectors \\
\begin{equation}
|\psi_{1}(k_{1})\rangle=
\left(
\begin{array}{c}
\cos\frac{\theta_{k_{1}}}{2}|k_{1}\rangle\\
\sin\frac{\theta_{k_{1}}}{2}|k_{1}+l_{1}\rangle\\
0
\end{array}
\right),~
|\psi_{2}(k_{1})\rangle=
\left(
\begin{array}{c}
-\sin\frac{\theta_{k_{1}}}{2}|k_{1}\rangle\\
\cos\frac{\theta_{k_{1}}}{2}|k_{1}+l_{1}\rangle\\
0
\end{array}
\right),~|\psi_{3}(k_{2})\rangle=
\left(
\begin{array}{c}
0\\
|k_{2}\rangle\\
0
\end{array}
\right),
\end{equation}\\
are the eigenkets of $H$, where $k_{1}<l_{2}-l_{1},~k_{2}<l_{1}$, 
$\theta_{k_{1}}=\tan^{-1}\left[2g_{l_{1}}(k_{1})/
(\varepsilon_{1}-\varepsilon_{2}-\omega l_{1})\right]$, 
and $g_{l_{1}}(k_{1})=\rho_{1}(k_{1})
\left[(k_{1}+l_{1})!/k_{1}!\right]^{1/2}$. 
The eigenvalues corresponding to 
above eigenkets are obtained to be\\
\begin{equation}\left\{
\begin{array}{l}
E_{1}^{0}(k_{1})=\frac{1}{2}(\varepsilon_{1}+\varepsilon_{2}+
\omega l_{1})+\omega k_{1}+ \sqrt{\frac{1}{4}(\varepsilon_{1}-
\varepsilon_{2}-
\omega l_{1})^{2}+g_{l_{1}}(k_{1})^{2}},
 \\[.15in]
E_{2}^{0}(k_{1})=\frac{1}{2}(\varepsilon_{1}+\varepsilon_{2}+
\omega l_{1})+\omega k_{1}-\sqrt{\frac{1}{4}(\varepsilon_{1}-
\varepsilon_{2}-
\omega l_{1})^{2}+g_{l_{1}}(k_{1})^{2}},
\\[.15in]
E_{3}^{0}(k_{2})=\varepsilon_{2}+\omega k_{2}.
\end{array}
\right.
\end{equation}\\
On the other hand, one can check that each element in the Kernel 
can be written as a linear combination of eigenkets in Eq. (141).
These results indicate that $U$ can be applied to $H$ 
to obtain the its remaining subspace. A simple calculation gives \\
\begin{equation}
H'=UHU^{\dagger}=\left (
\begin{array}{ccc}
\varepsilon_{1}+\omega(N+l_{2}-l_{1}), & g_{l_{1}}(N+l_{2}-l_{1}), 
& 0\\[.15in]
g_{l_{1}}(N+l_{2}-l_{1}), & \varepsilon_{2}+\omega(N+l_{2}), 
& g_{l_{2}}(N)\\[.15in]
0,&  g_{l_{2}}(N), & \varepsilon_{3}+\omega N
\end{array}
\right ).
\end{equation}\\
The Hamiltonian now becomes a matrix function of photon number 
$N$.
 One can follow
the procedure in the above subsections to obtain the 
eigenvalues and eigenkets of $H'$.
We omit these  here.

We now look at a system of three-level atoms with 
$\equiv$-configuration energy level, 
which interact with a one-mode cavity
as shown in Fig. 2.c. The Hamiltonian is 
\begin{equation}
H=\sum\limits_{i=1}^{3} \varepsilon_{i} S_{ii}+\omega a^{\dagger}a
+[\rho_{1}(N)a^{l_{1}}S_{12}+\rho_{2}(N)a^{l_{2}}S_{23}+h.c.],
\end{equation}
To solve the Hamiltonian, we introduce an operator matrix
 $V$ as
\begin{equation}
V=\left (
\begin{array}{ccc}
1 & 0 & 0\\
0 & F_{l_{1}} & 0\\
0 & 0 & F_{l_{1}+l_{2}}
\end{array}
\right ),
\end{equation}
$V$ evidently belongs to RUT. 
The Kernel of $V$ is\\
\begin{equation} 
K=\left\{|\psi_{1}^{0}(k_{1})\rangle=
\left(
\begin{array}{c}
0\\
|k_{1}\rangle\\
0
\end{array}
\right),~~|\psi_{2}^{0}( k_{2})\rangle=
\left(
\begin{array}{c}
0\\
0\\
|k_{2}\rangle
\end{array}
\right),~~~k_{1}\leq l_{1},~ k_{2}\leq l_{1}+l_{2} \right\}.
\end{equation}\\
Using the same method as in the above case of $V$-configuration, we 
obtain that this Kernel is isomorphic with to subset of the
eigenkets of $H$, where the eigenkets are \\
\begin{equation}
|\psi_{1}(k_{1})\rangle=
\left(
\begin{array}{c}
0\\
\cos\frac{\theta_{k_{1}}}{2}|k_{1}\rangle\\
\sin\frac{\theta_{k_{1}}}{2}|k_{1}+l_{2}\rangle
\end{array}
\right),~
|\psi_{2}(k_{1})\rangle=
\left(
\begin{array}{c}
0\\
-\sin\frac{\theta_{k_{1}}}{2}|k_{1}\rangle\\
\cos\frac{\theta_{k_{1}}}{2}|k_{1}+l_{2}\rangle
\end{array}
\right),~|\psi_{1}(k_{2})\rangle=
\left(
\begin{array}{c}
0\\
0\\
|k_{2}\rangle
\end{array}
\right),
\end{equation}\\
where $k_{1}<l_{1}$,~$k_{2}<l_{2}$, and $\theta_{k_{1}}=
\tan^{-1}\left[2g_{l_{2}}(k_{1})/
(\varepsilon_{2}-\varepsilon_{3}-\omega l_{2})\right]$.
The eigenvalues corresponding to above eigenkets are\\
\begin{equation}\left\{
\begin{array}{l}
E_{1}^{0}(k_{1})=\frac{1}{2}(\varepsilon_{2}+\varepsilon_{3}+
\omega l_{2})+\omega k_{1}+\sqrt{\frac{1}{4}(\varepsilon_{2}-
\varepsilon_{3}
-\omega l_{2})^{2}+g_{l_{2}}(k_{1})^{2}},
 \\[.15in]
E_{2}^{0}(k_{1})=\frac{1}{2}(\varepsilon_{2}+\varepsilon_{3}+\omega 
l_{2})
+\omega k_{1}-\sqrt{\frac{1}{4}(\varepsilon_{2}-\varepsilon_{3}-
\omega l_{2})^{2}
+g_{l_{2}}(k_{1})^{2}},
\\[.15in]
E_{3}^{0}(k_{2})=\varepsilon_{3}+\omega k_{2}.
\end{array}
\right.
\end{equation}\\
One can prove that an arbitrary element in the Kernel $K$ can
be  expressed by the above $|\psi_{1}(k)\rangle,
~|\psi_{2}(k)\rangle$ and $|\psi_{3}(k)\rangle$ in linear form.
This means that $K$ is isomorphic with one set of the eigenkets
of $H$. Therefore, the operator $U$ satisfies  Theorem IV, 
and can be applied to the remaining eigenket set of $H$. 
We obtain\\
\begin{equation}
H'=VHV^{\dagger}=\left (
\begin{array}{ccc}
\varepsilon_{1}+\omega N, & g_{l_{1}}(N), & 0\\[.15in]
g_{l_{1}}(N), & \varepsilon_{2}+\omega(N+l_{1}), & 
g_{l_{2}}(N+l_{1})\\[.15in]
0,&  g_{l_{2}}(N+l_{1}), & \varepsilon_{3}+\omega (N+l_{1}+l_{2})
\end{array}
\right ).
\end{equation}\\
$H'$ is a matrix, whose matrix elements are functions of photon
number $N$.  $H'$ can be solved by the regular way. We omit these 
here.

The above approach shows that all the $\Lambda,~V$ and 
$\equiv$-configuration atoms interacting with quantized cavity 
fields can be treated unitedly
by RUT method, where the right-unitary operators are simply  
diagonal matrices constructed by phaser operators.

In the last part of this section,  we would like
to show how to apply the RUT method to the atom-radiation 
interaction
system where the atoms have higher ($>3$) energy level. 
For an example, we here choose a simple case: 
four-level atoms with configuration  shown in Fig. 3.
Under the rotating-wave approximation,
the generalized Hamiltonian of the atoms interacting with
one-mode cavity field is
\begin{equation}
H=\sum\limits_{i=1}^{4} \varepsilon_{i} S_{ii}+\omega a^{\dagger}a
+[\rho_{1}a^{l_{1}}S_{12}+\rho_{2}a^{l_{2}}S_{23}+
\rho_{3}a^{l_{1}}S_{34}+h.c.],
\end{equation}
for simplicity, we assume $\rho_{i} ~i=1,2,3$ be constants.
To solve this Hamiltonian, we construct a right unitary
operator as
\begin{equation}
V_{4}=\left (
\begin{array}{cccc}
1 & 0 & 0 & 0\\
0 & F_{l_{1}} & 0 & 0\\
0 & 0 & F_{l_{1}+l_{2}}& 0\\
0 & 0 & 0 & F_{l_{1}+l_{2}+l_{3}}
\end{array}
\right ),
\end{equation}\\
The Kernel of $V_{4}$ is\\
\begin{equation} 
K=\left\{|\psi_{1}^{0}(k_{1})\rangle=
\left(
\begin{array}{c}
0\\
|k_{1}\rangle\\
0\\
0
\end{array}
\right),~~|\psi_{2}^{0}( k_{2})\rangle=
\left(
\begin{array}{c}
0\\
0\\
|k_{2}\rangle\\
0
\end{array}
\right),~~|\psi_{3}^{0}( k_{3})\rangle=
\left(
\begin{array}{c}
0\\
0\\
0\\
|k_{3}\rangle
\end{array}
\right) \right\}.
\end{equation}\\
where $k_{1}\leq l_{1},~ k_{2}\leq l_{1}+l_{2},~
k_{3}\leq l_{1}+l_{2}+l_{3}$. Within these 
ket vectors, the four-level system is reduced 
into a three-level case, one can further check out
that these ket vectors are isomorphic with 
a subset of the eigenkets of $H$. Which means that we can
use $V_{4}$ as a unitary operator 
to the other subspace of $H$. Let $H'=V_{4} H V_{4}^{\dagger}$,
then\\
\begin{equation}
H'=\left (
\begin{array}{cccc}
\varepsilon_{1}+\omega N, & g_{1}(N), & 0, & 0\\[.15in]
g_{1}(N), & \varepsilon_{2}+\omega(N+l_{1}), & g_{2}(N) & 0\\[.15in]
0,&  g_{2}(N), & \varepsilon_{3}+\omega (N+l_{1}+l_{2}) & g_{3}(N)    
\\[.15in]
0,& 0,&  g_{3}(N), & \varepsilon_{4}+\omega (N+l_{1}+l_{2}+l_{3})
\end{array}
\right ),
\end{equation}\\
where 
$g_{1}(N)=\rho_{1}
[( N+l_{1})!/N!]^{\frac{1}{2}}, ~g_{2}(N)=\rho_{2} 
[( N+l_{1}+l_{2})!/(N+l_{1})!]^{\frac{1}{2}},~
g_{3}(N)=\rho_{2} 
[( N+l_{1}+l_{2}+l_{3})!/(N+l_{1}+l_{2})!]^{\frac{1}{2}}$.
For the ket with a fix photon number $N$,
$H'$ is simply a $4\times 4$ constant matrix. Thus, its 
eigenkets and eigenvalues are easily obtained.

One can further follow this method to solve other configurations
of the four-level systems. These are  omitted here.\\

\begin{center}
\section*{ IV.~ Conclusion}
\end{center}
In conclusion, we have developed
 the right-unitary transformation (RUT) theory,
and initially discussed its applications in physics.
The first part of this paper attributes to the theory. We found that
the state space of any operator $U\in$ RUT (precisely 
$W$) is composed of two independent 
part:  $\{~|\Psi^{0}\rangle~\}$ and $\{~|\Psi^{1}\rangle~\}$,
where  $\{~|\Psi^{0}\rangle~\}$ is called the $Kernel$ of $U$, which 
satisfies $U\{~|\Psi^{0}\rangle~\}=0$. On the other hand, 
in the subspace $\{~|\Psi^{1}\rangle~\}$, 
$U$ acts as a unitary operator.
The properties of RUT such as semigroup, Kernel, etc.,
were discussed through several theorems. Based on these properties,
we concluded a general way  on how to apply  the RUT  
to a physical system.  For 
a physical quantity such as the Hamiltonian $H$, suppose its 
eigenstates
as $S=\{~|\Phi_{i}\rangle~, i=1,2,\cdots \infty ~\}$, then
\begin{equation}
H=\sum\limits_{i=1}^{\infty}E_{i}|\Phi_{i}\rangle\langle\Phi_{i}|,
\end{equation}
$E_{i}$ are eigenvalues. If a subset of $S$: 
$S_{1}=\{~|\Phi_{i}\rangle~, i=1,2,\cdots s~\}$ is checked to be
isomorphic with the Kernel of $U$, then  the supplement set 
of $S_{1}$ is evidently isomorphic with the unitary subspace of
$U$. Therefore,
\begin{equation}
H'=UHU^{\dagger}=
\sum\limits_{i=s+1}^{\infty}E_{i}|\Phi'_{i}\rangle\langle\Phi'_{i}|,
\end{equation}
where $|\Phi'_{i}\rangle=U|\Phi_{i}\rangle$. The new state space
$S'=\{~|\Phi'_{i}\rangle~, i=s+1,\cdots \infty ~\}$ is proved to
be complete. Equation (156) shows that the new frame $H'$
has the same spectrum (precisely a subset of the spectrum) as
$H$.

Based on the above results, in the second part 
of this paper we used the RUT method to  deal with
the systems of many-level atoms interacting 
with the quantized radiation fields,
where the RUT are the matrices constructed 
by the well-known phase operators. 
 We have studied two typical systems:  One is about the 
Jaynes-Cummings models, which were found to exhibit
some interesting effects, and have obtained much study 
in recent years. We solved a nonstationary generalized JC model, 
and found that atomic inversion of the system can be 
controlled through the 
external source. Another system  carefully studied 
is about the interaction of the three-level atoms with one or 
two-mode cavity  field. This system has been widely applied 
in various contexts of quantum optics such as lasing without 
inversion,
electromagnetically induced transparency, etc.
This paper provides a unified method for these topics.

We would like to point out that the RUT method can be applied
to some simplified quantum electrodynamic (QED) system, such as
the photon-electron, and phonon-electron interactions.
It will be discussed in the forthcoming presentation.\\

\acknowledgments
The author is deeply grateful to Dr.
 D. Finkelstein for his encouragement and support. 
The author is also indebted to Dr. B. Kennedy, F. T. Smith, S. Yu, C.L. 
Li, 
B. Berhane and I. Kulikov for their helpful discussions. 
The material is based upon research  supported in part by the
M\&H  Ferst Foundation, and by the NSF,  Grant No. PHY9211036.\\

\begin{center}

\appendix
\section*{}
\end{center}
For the initial state Eq. (63), 
\begin{equation}
\mu_{k}=\nu_{k}=e^{-|z|^{2}/2}\frac{z^{k}}{\sqrt{2k!}}\equiv f(z).
\end{equation}
Then,
\begin{equation}
\alpha_{k}^{+}=\left(
\frac{1-h_{k}}{1+h_{k}^{2}}\right) f(z),~~~
\alpha_{k}^{-}=\left(
\frac{h_{k}+h_{k}^{2}}{1+h_{k}^{2}}\right) f(z).
\end{equation}
From Eq. (51), we have  
\begin{equation}
h_{k}=\frac{\omega^{+}_{k}+\omega 
k+\frac{\omega_{0}}{2}}{g_{l}(k)}=
\frac{-g_{l}(k)}{\omega^{-}_{k}+\omega k+\frac{\omega_{0}}{2}},
\end{equation}
and 
\begin{equation}
h^{2}_{k}=-\frac{\omega^{+}_{k}+\omega k+\frac{\omega_{0}}{2}}
{\omega^{-}_{k}+\omega k+\frac{\omega_{0}}{2}}.
\end{equation}
Using these results, we rearrange $\alpha_{k}^{\pm}$ as
\begin{equation}
\alpha_{k}^{+}=\left(
\frac{\omega^{-}_{k}+\omega k+\frac{\omega_{0}}{2}+g_{l}(k)}
{\omega^{-}_{k}-\omega^{+}_{k}}\right) f(z),~~~
\alpha_{k}^{-}=-\left(
\frac{\omega^{+}_{k}+\omega k+\frac{\omega_{0}}{2}+g_{l}(k)}
{\omega^{-}_{k}-\omega^{+}_{k}}\right) f(z),
\end{equation}
with these expressions, we directly obtain Eqs. (64) and (65).

\begin{figure}
\caption{ (a)  shows that in the frame $H$, 
the transition from the atomic state 2 to state 1 is achieved by 
creating photons. (b) shows that in the transformed frame $H'$, 
there is no photon created or annihilated in the transition from 2 to 
1.}
\label{reduced}
\end{figure}

\begin{figure}
\caption{ (a) represents the transitions of 
the atoms with $\Lambda$ configuration energy levels. 
(b) indicates the $V$ configuration, and (c) 
the $\equiv$ configuration.}
\label{reduced}
\end{figure}

\begin{figure}
\caption{ Four-level atoms with ladder-configuration energies.}
\label{reduced}
\end{figure}


\begin{references}
\bibitem{time} Some nonunitary transformations such as $time$ $ 
reversal~$ $T$,  are not included in our consideration, 
since $T$ is an antilinear operator
without a proper definition of Hermitian conjugate.
\end{references}
\end{document}